\documentclass[aoas,preprint]{imsart}

\RequirePackage{amsthm,amsmath,amsfonts,amssymb,bbm}
\RequirePackage[authoryear]{natbib}
\RequirePackage[colorlinks,citecolor=blue,urlcolor=blue]{hyperref}
\RequirePackage{graphicx, soul}
\usepackage{hyperref}
\usepackage{longtable}
\usepackage{xcolor}
\usepackage{multirow}
\usepackage{url}
\usepackage{enumitem}
\usepackage{array}
\newcolumntype{L}{>{\arraybackslash}m{8cm}}
\startlocaldefs
\DeclareMathOperator*{\argmax}{argmax}

\endlocaldefs

\begin{document}

\begin{frontmatter}
\title{Bayesian decision theory for tree-based adaptive screening tests with an application to youth delinquency}
\runtitle{Bayesian decision theory for adaptive screening tests}

\begin{aug}
\author[A]{\fnms{Chelsea} \snm{Krantsevich}\ead[label=e1]{chelsea.krantsevich@asu.edu}},
\author[A]{\fnms{P. Richard} \snm{Hahn}\ead[label=e2,mark]{prhahn@asu.edu}}
\author[A]{\fnms{Yi} \snm{Zheng}\ead[label=e3,mark]{Yi.Isabel.Zheng@asu.edu}}
\and
\author[B]{\fnms{Charles} \snm{Katz}\ead[label=e4,mark]{CHARLES.KATZ@asu.edu}}

\address[A]{School of Mathematical and Statistical Sciences, Arizona State University, \printead{e1,e2,e3}}

\address[B]{School of Criminology and Criminal Justice, Arizona State University, \printead{e4}}
\end{aug}

\begin{abstract}

Crime prevention strategies based on early intervention depend on accurate risk assessment instruments for identifying high risk youth. It is important in this context that the instruments be convenient to administer, which means, in particular, that they should also be reasonably brief; adaptive screening tests are useful for this purpose. Adaptive tests constructed using classification and regression trees are becoming a popular alternative to traditional Item Response Theory (IRT) approaches for adaptive testing. However, tree-based adaptive tests lack a principled criterion for terminating the test. This paper develops a Bayesian decision theory framework for measuring the trade-off between brevity and accuracy, when considering tree-based adaptive screening tests of different lengths. We also present a novel method for designing tree-based adaptive tests, motivated by this framework. The framework and associated adaptive test method are demonstrated through an application to youth delinquency risk assessment in Honduras; it is shown that an adaptive test requiring a subject to answer fewer than 10 questions can identify high risk youth nearly as accurately as an unabridged survey containing 173 items.


\end{abstract}


\end{frontmatter}

\section{Introduction}
\label{section:Introduction}
Screening tests, or {\em instruments}, serve as an important aid to decision making in many fields (e.g. education, mental health, or social work) by sorting the population of test takers into groups which then receive different follow-up services (curricula, therapies, or support resources). Designing an effective screening instrument -- developing the test items and determining the order and manner in which they are administered -- presents many challenges. This paper looks narrowly at the trade-off between two competing virtues in a screening instrument: brevity and accuracy. Lengthy instruments can cause exam fatigue for participants as well as administrators, potentially limiting the number of individuals that can be screened at all. Conversely, a conveniently brief assessment with poor accuracy is equally as fruitless. The problem considered here is: starting with a screening test comprised of many questions (a large item bank), can a shorter screening test be derived without sacrificing accuracy relative to the full test? The basic statistical challenge in navigating this trade-off is that the accuracy of both the full-length test and the abridged test must be estimated from data. A secondary challenge is the computational search for subsets of items which preserve accuracy. This paper proposes solutions to both of these issues. Statistical uncertainty in the instrument accuracies is addressed by casting the design of the abridged instrument as a problem in Bayesian decision theory. The computational challenge is addressed by constructing the adaptive screening test as a decision tree, allowing us to adapt existing algorithms for the purpose of designing shortened instruments. \footnote{The code used for this paper is available at \url{https://github.com/chelsea-k/adaptive-tests/}}



Our motivating application is the design of a brief screening test to identify youth who are at high risk of falling into delinquent behavior, so that they may receive additional social support intended to mitigate that risk. Specifically, we consider data from Honduras, where decades of political, civil, and economic instability have made gang recruitment and violent crime a major concern (\cite{Meyer2019}, \cite{UNODC2018}). Certain targeted interventions, such as family counseling and community support resources, have demonstrated significant promise in reducing risk factors of criminal behavior for at-risk youth in Honduras (\cite{Katz2021}). In order to allocate these limited resources in an effective way, a screening instrument is deployed to identify youth with the highest risk of delinquency.

\section{Previous work}
\label{section:Previous work}
This paper brings together three distinct strands of research. First, we add to a well-established literature on youth risk assessment, specifically their application to crime prevention programs aimed at youth in Central America. In this context we reanalyze data from Honduras and show that an adaptive screening test consisting of only a handful of items can provide comparably accurate risk assessment to a questionnaire with over a hundred questions. Second, we build on recent work using classification trees to design abridged diagnostic tests. To this literature we add a principled approach to constructing the tree and determining its maximum depth, using ideas from Bayesian decision theory to evaluate the trade-offs between instruments of different lengths. Finally, this Bayesian decision theoretic approach is a natural extension of ideas developed in recent work on utility-based posterior summarization. Here, we apply these ideas in the novel context of adaptive screening tests.

\subsection{Youth risk assessment in Honduras and elsewhere}

For a broad overview of the difficulties facing youth in Honduras, please consult \cite{BerkSeligson2014b}. Here we focus on risk assessment tools used in crime prevention, which has recently gained momentum as an effective alternative to more aggressive suppression strategies.

As a key component of crime prevention, so-called ``secondary prevention'' programs identify individuals within high risk communities who are at an especially high risk for criminal activity, and provide them with targeted interventions. To effectively execute this secondary prevention strategy, high risk youth must first be identified via a screening tool and are subsequently enrolled in the intervention.

For the model utilized between 2013 and 2015 in Honduras specifically, high risk youth were first identified using a Spanish adaptation of the Youth Services Eligibility Tool (YSET) (\cite{Hennigan2014}), and then enrolled in a seven-module family counseling program. This model represented the first time empirical data was utilized for identifying the youth with the highest risk of criminal behaviors. Following initial successes, a more locally focused risk assessment tool was created, incorporating screening tools from around the world. The data for the present work consists of responses to this revised Honduran YSET, and is described more fully in Section \ref{section:Data}.

The risk assessment tools utilized in Honduras are based on a large body of research surrounding the risk factor paradigm. Risk factors are characteristics that increase the likelihood of a given problem behavior, whereas protective factors are ones that reduce this likelihood (\cite{Arthur2002}). These factors are typically categorized under domains such as community, family, school, peer, and individual (\cite{Howell2005}). See Table 1 in Section 2 of the Supplementary Material (\cite{Krantsevich2022}) for a list of risk and protective factors measured in our work. 

The risk factor paradigm entered the youth delinquency sphere in 1992 with \cite{Hawkins1992}, who provided a comprehensive review of the literature on risk and protective factors related to substance abuse in adolescents. In subsequent years, multiple groups developed youth risk assessment tools, including three that were used to expand the item bank for the revised Honduran YSET: the Communities That Care (CTC) Youth Survey (\cite{Arthur2002},  \cite{Arthur2007}),  the Eurogang Youth Survey (\cite{Weerman2009}), and the Youth Eligibility Services Tool (YSET) (\cite{Hennigan2014}, \cite{Hennigan2015}), a Los Angeles-specific adaptation of the empirically-developed Gang Risk of Entry Factors instrument. 


While these instruments have been deployed in countries around the world, they were largely developed for use in the United States and Europe. Research on youth risk assessments for secondary prevention programs within developing countries includes \cite{Katz2010} and \cite{Maguire2011}, focusing on the Caribbean nation of Trinidad and Tobago, and \cite{Webb2016}, focusing on the Northern Triangle nation of El Salvador. These works including protective factors in addition to risk factors, which provide an avenue to learn about positive interventions the community can undertake. For more information on risk and protective factors in low- and middle-income countries, we refer the reader to the systematic reviews of \cite{Murray2018} on risk and protective factors for antisocial behavior, and \cite{Higginson2018} on risk and protective factors related to gang membership. 

\subsection{Adaptive testing}
An {\em adaptive test} is one where the next question a subject encounters depends on her answer to the previous question or questions. Adaptive tests are a powerful approach for developing shortened risk assessments, because while any given test taker may only see a small number of questions, the wide variety of available questions allows different subjects to be classified more accurately than if every subject was administered the same small number of questions. Traditionally, adaptive tests have been constructed using item-response theory (IRT), which requires estimating the latent constructs of each test taker at the time of testing (\cite{Wainer2000}). Examples of IRT-based adaptive testing in the academic or personnel selection setting include the Graduate Management Admission Test (\cite{Rudner2010}), the Graduate Record Examination (\cite{Almond1998}), and the Armed Services Vocational Aptitude Battery (\cite{Sands1997}). 

The cornerstone of IRT is the Item Response Function (IRF), which describes, for each item in the item bank, how the participant's response to that item depends on their value of the latent trait being measured (delinquency risk). An adaptive test designed based on IRT proceeds by starting with an initial risk level, administering the most informative item (based on the participant's risk estimate and each item's IRF), updating the risk estimate based on their response, and iterating until a stopping criterion is satisfied. Deploying an IRT-based adaptive test requires several specifications: the IRF family (and calibrating individual item parameters), the algorithm for estimating the latent trait, the criterion for selecting successive items, and the stopping criterion (\cite{Chang2004}, \cite{Chang2015}, \cite{vanderLinden2008}, \cite{Wainer2000}). 

For a comprehensive overview on IRT, see \cite{vanderLinden1997}. We also refer interested readers to  \cite{Hambleton1991}, \cite{Embretson2000}, and \cite{deAyala2009} for more accessible treatments of the subject.

There are two major downsides to IRT-based adaptive tests in our application. One, the real-time estimation of the latent trait parameter requires intense computational resources, necessitating test administration via a laptop or computer and making the screening process more challenging. Two, as noted by \cite{Gibbons2016} and \cite{Zheng2020}, many constructs in practical screening or diagnostic contexts are multidimensional. While multidimensional extensions of IRT-based adaptive tests to a handful of dimensions exist (\cite{Frey2009}, \cite{Gibbons2016}, \cite{Paap2017}, \cite{Wang2011}, \cite{Wang2012}, \cite{Yao2014}), this is likely unsuitable for capturing the relationship between the 38 different scales in our risk assessment. 


\subsubsection{Adaptive testing using classification trees}
\label{section:CART}

Tree-based adaptive tests, constructed entirely beforehand using classification trees, are a recently explored alternative to IRT-based tests, and have already been used for measuring a variety of medical and behavioral constructs. \cite{Zheng2020} is the first tree-based adaptive test to our knowledge to be used for assessing youth risk of delinquency; the authors utilized item-response data from crime prevention programs in Honduras, comparing the performance of several tree-based adaptive tests fit using the CART algorithm (\cite{Breiman1984}), including one fit to synthetic data generated using the Synthetic Minority Over-sampling Technique (SMOTE, \cite{Chawla2002}).

The tree-based approach to adaptive testing involves collecting responses to a large number of items, as well as a true outcome measurement; a classification tree is then fit to this data to maximize predictive accuracy. Specifically, the Classification And Regression Trees (CART) algorithm, introduced by \cite{Breiman1984}, is applied to the item response-outcome data. Here we review the basics for reference. A modern survey of CART and other tree-growing methods can be found in \cite{Loh2011}.

A classification or regression tree $T$ partitions a covariate space $\mathcal{X}$ into $k$ disjoint hypercubes, $A_1$, $A_2$, \dots, $A_k$, by repeatedly splitting $\mathcal{X}$ one variable at a time. Each internal node of a final fitted tree contains a splitting variable and an associated cutpoint, $x_i \leq b$. The number of leaf nodes $k$ corresponds to the size of the partition, and the data stored in each leaf node of the fitted tree represents the output of the tree function. In a regression tree, values $\mu_1, \dots, \mu_k\in \mathbb{R}$ are associated to the $k$ leaf nodes, so that $\mathrm{x}\in A_j$ implies $T(\mathrm{x})=\mu_j$, $1\leq j \leq k$. In a classification tree with $c$ classes, the $j^{th}$ leaf node contains a probability distribution $\{p_{1j}, p_{2j}, \dots, p_{cj}\}$ over the classes, and for $\mathrm{x}\in A_j$, $T(\mathrm{x})$ is the class with the highest probability: $T(\mathrm{x})=\text{argmax}_{i \in \{1, \dots, c\}} p_{ij}$.

In the classic CART algorithm, the tree is fit to data with the goal of minimizing node impurity according to a given criterion (e.g. mean squared error for regression and Gini index for classification). The algorithm proceeds by first growing a very deep tree, then pruning back to the final tree. See \cite{Breiman1984} for details on the CART growing and pruning algorithms.

\begin{figure}
    \centering
    \includegraphics[width = 0.6\linewidth]{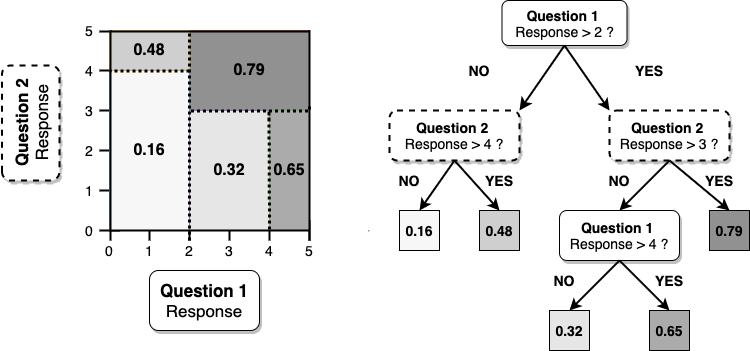}
    \caption{A tree-based adaptive test. The splitting variables are $\mathcal{X} = \{\text{Question } 1, \text{ Question }2\}$. A subject with responses of 3 and 4, respectively, would land in the right-most leaf node and have ``at-risk'' probability 79\%.}
    \label{fig:CART}
\end{figure}

To use a tree as an adaptive test, items are used as splitting variables and item responses as cutpoints. After fitting the classification tree to item response--risk class data, a new subject takes the tree-based adaptive test by first answering the root node item, then moving right or left according to their response and the cutpoint. Subsequent items are administered based the pattern of item responses. The assessment ends when the subject lands in a terminal or ``leaf'' node, with their predicted risk class being the one assigned to that leaf node. Alternatively, one can use the ``at-risk'' probability stored in the leaf node as the tree output, and assign a risk class of ``at-risk'' to youth above a certain probability, with this cutoff determined separately. See Figure \ref{fig:CART} for an example of the latter approach, with two items.

In the past several years, multiple groups have experimented with tree-based adaptive tests in various settings, including measuring quality of life in Multiple Sclerosis patients (\cite{Michel2018}), predicting risk of suicide attempt (\cite{DelgadoGomez2016}) and reproducing a clinician's diagnosis of depression (\cite{Gibbons2013}). 

\cite{Gibbons2013} depart from traditional tree-growing approaches by fitting the tree to a large amount of artificial data, generated as follows (\cite{Gibbons2019}): first, item response vectors are created via local perturbations\footnote{By ``local perturbation'' we mean that item response vectors were selected uniformly at random from vectors that are in a neighborhood of the observed item response vectors in terms of the $L_1$ distance.}. Next, a Random Forest model (\cite{Breiman2001}) is fit to the original data, and used to predict artificial outcome classes for the artificial item response vectors. A single classification tree is then fit to this large artificial dataset and the fitted tree is used as the adaptive test. \cite{Gibbons2013} and \cite{Gibbons2016} claim that the use of artificial data increases stability of the final classification tree, although they do not discuss details. Our method, while also utilizing artificially generated data, does so for fundamentally different reasons rooted in past work on posterior summarization for model selection (see Section \ref{section:Utility-based posterior summarization}). Synthetic data in our case is used to approximate a posterior distribution of a utility function, and we select the optimal decision tree according to this utility. Further details are discussed in Section \ref{section:The_adaptive_screening_decision_problem}.

While tree-based adaptive tests have several advantages over IRT, including ease of deployment and fewer modeling assumptions, there is no clear standard to determine how deep to grow the tree, or in other words, when to terminate the test; instead, this choice is made by the default regularization parameters in the tree-growing software. The exam length has important implications; in particular, shortening the exam too much can lead to unacceptable levels for instrument sensitivity and specificity. However, to the best of our knowledge there is no standard stopping criterion for a tree-based adaptive test to ensure a certain sensitivity and specificity.

\subsection{Utility-based posterior summarization}
\label{section:Utility-based posterior summarization}
A recent line of research has recast the problem of variable or model selection as one of posterior summarization. The idea is to find a single model-summarizing ``action'' that minimizes a penalized loss function which favors parsimony. In the present context, the idea is to find a shortened screening test that is suitably accurate relative to the non-shortened instrument. This line of work began with  \cite{Hahn2015} for linear regression models and has been expanded in various directions subsequently \citep{Bashir2019, Puelz2017, Woody2019}.  The technique explored in these papers is a two stage process: first, a highly flexible and accurate model is fit; then, draws from the posterior distribution are projected onto simpler structures, producing low-dimensional model summaries. In this way, an analyst may visualize how much accuracy (however that is defined) relative to an ``ideal'' non-simplified model. In the risk assessment setting studied here, the ``ideal'' non-simplified model is a non-shortened screening instrument that incorporates responses to every item in the item bank in order to predict the probability of exhibiting delinquent behavior. In this step, we may use any state-of-the-art predictive algorithm to obtain an ``at-risk'' probability estimate. Then, we consider the trade-off in model accuracy that is made by administering a greatly-shortened adaptive test, in which each subject sees only a small number out of the many items available. We use a Bayesian decision theory framework to formalize these trade-offs.

\section{A decision theory perspective on adaptive screening}
\label{section: Adaptive screening as decision theory} 

In the following sections, we recall general elements of Bayesian decision theory, then explain its particular application in risk assessment of youth delinquency.

Throughout the paper, we use calligraphy $\mathcal{X}$ and $\mathcal{Y}$ to denote the support of item response vectors and risk classes, upper-case $\mathrm{X}$ and $Y$ to denote a random vector/variable representing an item response vector or risk class, and lower-case $\mathrm{x}$ and $y$ to denote a single instantiation of the random vector/variable. As is standard in Bayesian statistics, we treat model parameters as random variables, rather than fixed parameters of the data generating process of $\mathrm{X}$ and $Y$; the random vector of all unknown model parameters is represented by $\theta$, with $\Theta$ being its support and $\theta^{(j)}$ a single instantiation. When referring to observed data, we use subscripts $\mathrm{x}_{1:n}$ and $y_{1:n}$; synthetic data are denoted by $\tilde{\mathrm{x}}$ and $\tilde{y}$. 


\subsection{Review of Bayesian decision theory}
\label{section:Review_of_Bayesian_decision_theory}

In this section we provide an introduction to Bayesian decision theory using the terminology of \cite{Parmigiani2010}, according to which an analyst chooses from among a \emph{set of actions}, $\Gamma$. Each action $\gamma:\mathcal{X}\rightarrow\{0,1\}$ has consequences that depend on an unknown \emph{state of the world}, $y\in \mathcal{Y}$. In order to evaluate the merits of possible actions, a quantitative value is assigned to each possible (action, state) pair, either a \emph{utility} value $U(\gamma(\mathrm{x}), y)$ or a \emph{loss} value $L(\gamma(\mathrm{x}), y)$. With the \emph{utility function} framework, which we employ, the analyst chooses the action that maximizes (in some sense) a utility.

We adopt the \emph{expected utility principle}, which implies the chosen action maximizes expected utility over a target population with density $f(\mathrm{x},y)$. This expected utility is
\begin{equation}
\mathbb{E}U(\gamma) := \mathbb{E}[U(\gamma(\mathrm{X}),Y))] =  \int_{\mathcal{X}} \int_{\mathcal{Y}} U(\gamma(\mathrm{x}),y)f(\mathrm{x},y)\; dy \; d\mathrm{x},
\end{equation}

and the optimal action is
\begin{align*}
\gamma^* = \text{argmax}_{\gamma\in\Gamma}\mathbb{E}U(\gamma).
\end{align*}

To summarize, our decision theory formulation consists of:
\begin{quote}
\normalsize
\begin{enumerate}
\item A \emph{utility function} $U.$
\item A \emph{target population} defined by a distribution function $F_{\mathrm{X},Y}$.
\item A \emph{set of actions}, denoted $\Gamma.$
\end{enumerate}
\end{quote}
These three elements come together in defining our expected (integrated) utility $\mathbb{E}U(\gamma)=\mathbb{E}[U(\gamma(\mathrm{X}),Y))]$, where $\gamma \in \Gamma$ and $\mathbb{E}(\cdot)$ denotes expectation with respect to $F_{\mathrm{X},Y}$.

In our application to youth risk assessment, an action $\gamma$ is a tree-based adaptive screening test, which takes the youth's item responses $\mathrm{x}\in \mathcal{X}$, and assigns an outcome of either ``at-risk'' ($\gamma(\mathrm{x})=1$) or ``not at-risk'' ($\gamma(\mathrm{x})=0$), determining enrollment into a secondary prevention program. We apply the preceding framework to the youth delinquency problem as follows:
\begin{quote}
\normalsize
\begin{enumerate}
\item Our \emph{utility function}, $U$, is a weighted average of sensitivity and specificity.
 \item Our \emph{target population} is the group of youth to be screened for risk of problem behaviors. We let $f(\mathrm{x},y)$ denote the joint density function of item responses and risk status for youth in the target population.
 \item  Our \emph{set of actions}, $\Gamma$, is a collection of candidate screening tests of varying lengths. (This action space will be populated using a tree growing algorithm, detailed later.) 
\end{enumerate}
\end{quote}  
Sections  \ref{section:Specifying_the_utility_function},  \ref{section:Specifying_the_target_population} and \ref{section:Populating_the_action_space}, respectively, describe these steps in greater detail.

In practice, the density function $f(\mathrm{x}, y)$ is unknown and must be estimated from available data. To do so, we will parametrize $f$ by a vector $\theta$, which we will estimate via Bayesian inference. We choose a prior $\pi(\theta)$ and, after conditioning on data $(\mathrm{x}_{1:n}, y_{1:n})$, arrive at a posterior $\pi(\theta\mid \mathrm{x}_{1:n}, y_{1:n})$. Rather than integrating over the estimation uncertainty in $\theta$ as would be done in traditional Bayesian decision theory, we will instead consider posterior uncertainty of the utility $\mathbb{E}U(\gamma, \theta)$, defined as
\begin{equation}
\label{eqn: conditional utility}
\mathbb{E}U(\gamma, \theta) :=  \int_{\mathcal{X}} \int_{\mathcal{Y}} U(\gamma(\mathrm{x}),y)f(\mathrm{x}, y \mid \theta) \; dy \; d\mathrm{x}.
\end{equation}
 As a function of $\theta$, $\mathbb{E}U(\gamma, \theta)$ is itself a random variable, which we denote $\mathbb{E}U_\theta(\gamma)$ for notational convenience. In this paper, we will be interested in the posterior distribution of $\mathbb{E}U_\theta(\gamma)$ induced by the posterior distribution over $\theta$.

\subsection{The adaptive screening decision problem}
\label{section:The_adaptive_screening_decision_problem}

Here we describe how the three steps of the Bayesian decision theory framework are applied to adaptive screening tests for youth delinquency. Recall that our set of actions $\Gamma$ is comprised of adaptive screening tests $\gamma$ for assessing risk of youth delinquency. Each test $\gamma$ consists of two parts:
\begin{quote}
\normalsize
\begin{enumerate}
\item A binary tree $T:\mathcal{X}\rightarrow (0,1)$ representing the screening test (see Figure \ref{fig:CART}, right). $T$ predicts the ``at-risk' probability $T(\mathrm{x})$, given item responses $\mathrm{x}\in \mathcal{X}$. 
\item A threshold function $\text{Thr}_C:(0,1)\rightarrow\{0,1\}$ that maps the probability $T(\mathrm{x})$ to a risk status prediction via a cutoff $C\in [0,1]$:
\begin{eqnarray*}
\text{Thr}_C(T(\mathrm{x})) &= \begin{cases} 0, \text{``not at-risk''} &\mbox{if } T(\mathrm{x}) < C \\
1, \text{``at-risk''} & \mbox{if } T(\mathrm{x}) \geq C \end{cases}
\end{eqnarray*}
\end{enumerate}
\end{quote}
Put together, the adpative test is $\gamma(\cdot) = \text{Thr}_C(T(\cdot))$, where $\gamma(\mathrm{x})\in \{0,1\}$ for any given set of item responses $\mathrm{x}$. The framework described in the next three sections provides a way to compare different youth risk assessments of this form. 

\subsubsection{Specifying the utility function}
\label{section:Specifying_the_utility_function}

Step 1 of the framework is specifying a utility function $U$. The adaptive test $\gamma$ should maximize $\mathbb{E}U_
\theta(\gamma)$, the expectation of $U$ with respect to the density $f(\mathrm{x},y)$ (which is parameterized by $\theta$) over our target population. 

\begin{figure}
    \centering
    \begin{minipage}{0.49\textwidth}
        \centering
       \includegraphics[width = 0.8\textwidth]{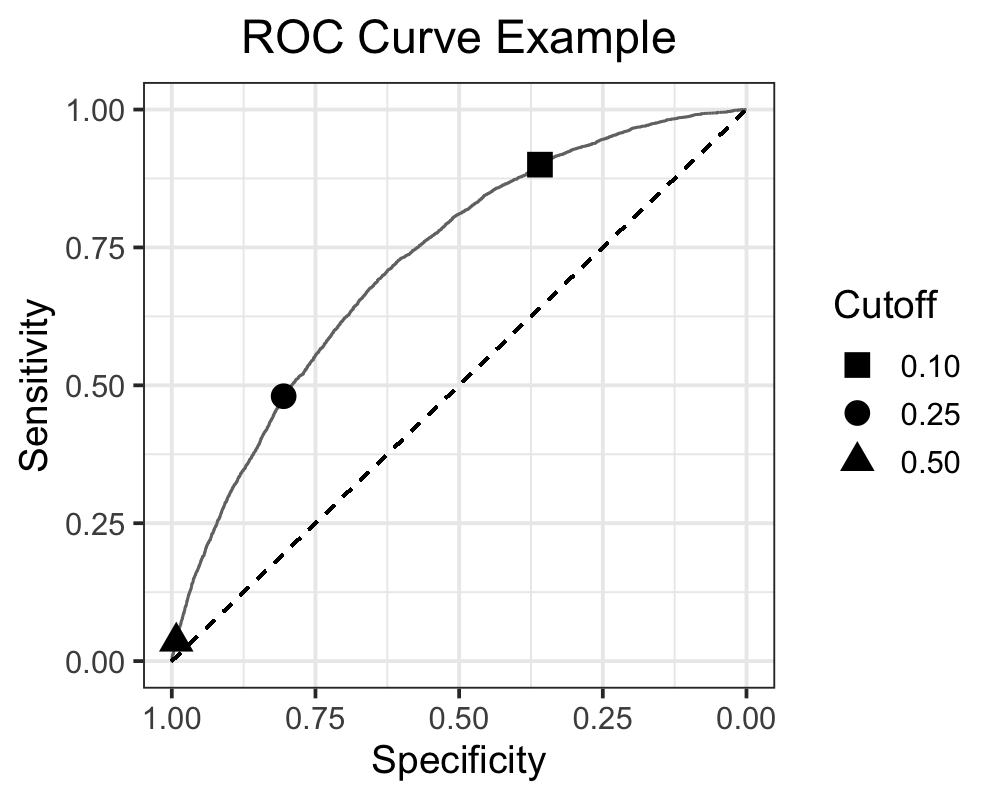}
    \caption{Each cutoff $C$ results in a particular (Specificity, Sensitivity) point on the ROC curve. Lowering the cutoff increases sensitivity and decreases specificity. Choosing a cutoff means choosing an acceptable (Specificity, Sensitivity) combination, or point on the ROC curve.}
    \label{fig:ROC_example}
    \end{minipage}\hfill
    \begin{minipage}{0.49\textwidth}
        \centering
      \includegraphics[width = 0.62\textwidth]{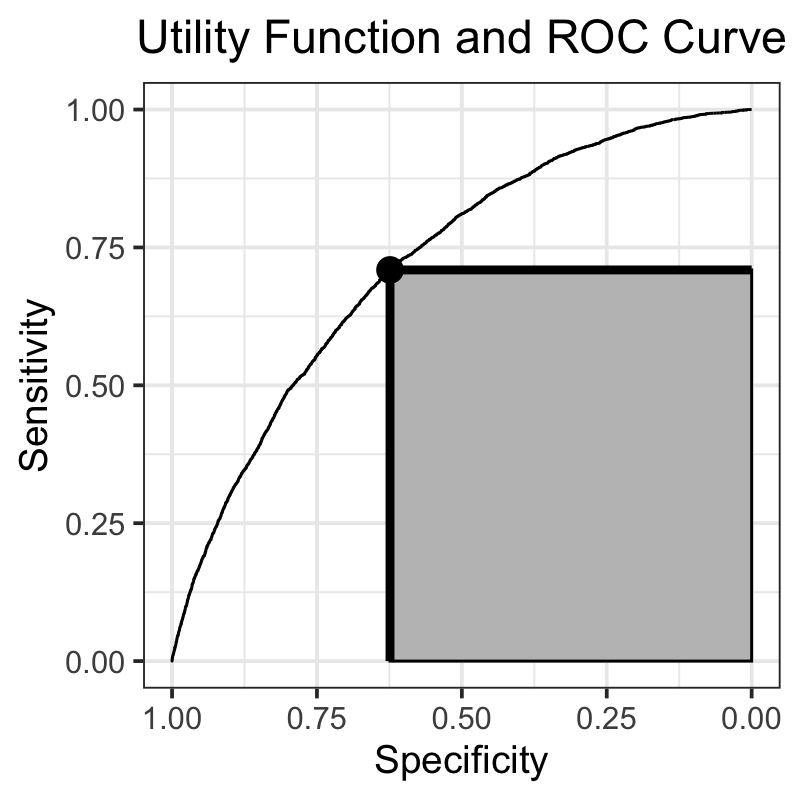}
    \caption{The cutoff that maximizes the utility function (\ref{eqn:expected_utility_intuitive}) for $w=0.5$ is the point on the ROC curve that maximizes the perimeter of the shaded rectangle. The height and width of the rectangle are sensitivity and specificity, respectively, for that cutoff.}
    \label{fig:Utility_on_ROC}
    \end{minipage}
\end{figure}

In our application, we want the utility function to carry practical significance for the adaptive test. Two important quantities are sensitivity and specificity, which measure the true positive rate and true negative rate, respectively:
\begin{align*}
\text{Sensitivity} = \text{Pr}(\gamma(\mathrm{X})=1\mid Y=1), \hspace{0.2cm} \text{Specificity} = \text{Pr}(\gamma(\mathrm{X})=0\mid Y=0).
\end{align*} 
 As a reminder, $\gamma$ is an adaptive test mapping item responses $\mathrm{X}$ to a risk status class $Y$ (``at-risk'' means $Y=1$, ``not at-risk'' means $Y=0$). Ideally, sensitivity and specificity would both be $1$. In practice, there is a trade-off between these two quantities, based on the cutoff $C$. A high cutoff means that many predicted probabilities will be below the threshold and consequently labeled ``not at-risk'', leading to high specificity and low sensitivity. A low cutoff leads to more ``at-risk'' class predictions, increasing sensitivity and reducing specificity. This trade-off can be visualized in a Receiver Operating Characteristic (ROC) curve, shown in Figure \ref{fig:ROC_example}. 

To incorporate the importance of both sensitivity and specificity, our expected utility $\mathbb{E}U_\theta(\gamma)$ is equal to a weighted average of the two, for a user selected weight $w\in (0,1)$:
\begin{align}
\label{eqn:expected_utility_intuitive}
\mathbb{E}U_\theta(\gamma)=w\cdot \text{Sensitivity}(\gamma) + (1-w)\cdot \text{Specificity}(\gamma).
\end{align}
See Section 1 of the Supplementary Material (\cite{Krantsevich2022}) for the point-wise (individual) specification of the utility function $U$ which induces this expected (population level) utility. For $w=0.5$, this utility function can be directly visualized within a ROC curve as shown in Figure \ref{fig:Utility_on_ROC}. 

Since the final adaptive tree and the associated sensitivity and specificity highly depend on $w$, we recommend carefully selecting the value of the weight in conjunction with stakeholders who understand the implications of favoring sensitivity or specificity for the population where test will ultimately be deployed. Multiple values of $w$ can and should be examined via the methods presented in Section \ref{section:Results}. 

With this utility function and a value of $w$ specified, the optimal action $\gamma^*$ is the tree-based adaptive test (i.e., ``at-risk'' probability prediction and associated cutoff) that maximizes this weighted average. 

Since the expected utility of a given action $\gamma$ is a simple expression at the population level, we can evaluate $\mathbb{E}U_\theta(\gamma)$ over a sample from the target population by directly computing sensitivity and specificity of $\gamma$ for a particular set of item responses and true risk classes. To be more specific, after drawing a sample $\{\tilde{\mathrm{x}}_{ij}, \tilde{y}_{ij}\mid \theta^{(j)}\}_{i=1}^N$ from the target population (where $\tilde{\mathrm{x}}_{ij}$ is an item response vector, $\tilde{y}_{ij}$ is the risk class, and $\theta^{(j)}$ is a single fixed draw from the posterior $\pi(\theta \mid \mathrm{x}_{1:n}, y_{1:n})$---see Section \ref{section:Specifying_the_target_population}), we compute a draw $\mathbb{E}U_{\theta^{(j)}}(\gamma)$ as
\begin{equation}
\label{eqn:computing_utility_from_sample}
\mathbb{E}U_{\theta^{(j)}}(\gamma)= w\cdot \frac{\sum_{i=1}^N\mathbbm{1}(\gamma(\tilde{\mathrm{x}}_{ij})=1, \tilde{y}_{ij}=1)}{\sum_{i=1}^N\mathbbm{1}( \tilde{y}_{ij}=1)} + (1-w)\cdot \frac{\sum_{i=1}^N\mathbbm{1}(\gamma(\tilde{\mathrm{x}}_{ij})=0, \tilde{y}_{ij}=0)}{\sum_{i=1}^N\mathbbm{1}( \tilde{y}_{ij}=0)}.
\end{equation}

In the next section, we describe how to sample from the target population in order to obtain draws of $\mathbb{E}U_\theta(\gamma)$ for any given action $\gamma$.

\subsubsection{Specifying the target population}
\label{section:Specifying_the_target_population}

Step 2 of the Bayesian decision theory framework is specifying a target population over which we seek to maximize $\mathbb{E}U_\theta(\gamma)$. In our application, that means defining a specific group of youth for which we want to optimize our adaptive test. After specifying the target population, the optimal action $\gamma^*$ (the ``optimal'' adaptive test) would maximize the weighted average of sensitivity and specificity for this group specifically. The target population can be all Honduran youth, or more specific to youth of a certain age, living in a particular neighborhood, and so forth.

After specifying the target population, we draw synthetic samples from the joint density $f(\mathrm{x}, y)$ of the item responses $\mathrm{X}$ and risk class $Y$ in the target population. We use the composite model specification
\begin{equation*}
f(\mathrm{x},y)=f(\mathrm{x})f(y\mid \mathrm{x})
\end{equation*}  
and specify the random variable $\theta$ parameterizing $f(\mathrm{x},y)$ as 
\begin{equation*}
\theta=(\theta_{\mathrm{X}}, \theta_Y),
\end{equation*}
with $\theta_{\mathrm{X}}$ parameterizing $f(\mathrm{x})$ and $\theta_Y$ parameterizing $f(y\mid \mathrm{x})$. This specification allows for additional flexibility in modeling the relationship between the item responses $\mathrm{X}$ and the risk of delinquency $Y$. 
Practically, we draw synthetic data from $f(\mathrm{x}, y)$ as follows: 
\begin{quote}
\normalsize
\begin{enumerate}
\item Fit each component of the composite form using a Bayesian model: one for $f(\mathrm{x})$ with unknown parameters $\theta_{\mathrm{X}}$, and one model for $f(y\mid \mathrm{x})$ with unknown parameters $\theta_Y$. 
\item For each posterior draw $\theta^{(j)}=(\theta^{(j)}_\mathrm{X}, \theta^{(j)}_Y)$, $1\leq j \leq D$, draw samples 
\begin{equation*}
\left\{\tilde{\mathrm{x}}_{ij}, \tilde{p}_{ij}, \tilde{y}_{ij}\mid \theta^{(j)}\right\}_{i=1}^N
\end{equation*}
 from the conditional predictive distribution $f(\tilde{\mathrm{x}}, \tilde{y}\mid \theta^{(j)})$. Here, $\tilde{\mathrm{x}}_{ij}$ are the synthetic item responses, $ \tilde{p}_{ij} = \mathbb{E}(\tilde{Y}\mid \tilde{\mathrm{x}}_{ij}, \theta_Y^{(j)})$ is the synthetic probability of belonging to class $Y=1$, and  $\tilde{y}_{ij}$ is the synthetic class status.
\end{enumerate}
\end{quote}
Taken together, we will have a sample of size $N$ for each posterior draw $\theta^{(j)}$, $1\leq j \leq D$, which is $N\cdot D = M$ synthetic data in total; this data is denoted $\{\tilde{\mathrm{x}}_k, \tilde{p}_{k}, \tilde{y}_k\}$, $1\leq k \leq M$. 

Since we fit two models corresponding to different components of the same composite model specification, we use a single dataset for fitting the models for $f(\mathrm{x})$ and $f(y\mid \mathrm{x})$. Modeling details are provided in Section \ref{section:Modeling}; more specifics on sampling can be found in Appendix \ref{Appendix:Integrating}. As a reminder, the ``synthetic data'' in this setting is merely a computational approach for evaluating the integrals at the heart of the decision theory framework.

\subsubsection{Populating the action space}
\label{section:Populating_the_action_space}

In this section we describe Step 3 of the framework, populating the action space $\Gamma$. In our application, $\Gamma$ consists of tree-based adaptive tests; each $\gamma\in \Gamma$ is of the form  form $\gamma (\cdot) = \text{Thr}_{C_T}(T(\cdot))$, where $T$ is a binary regression tree and $C_T$ is the cutoff for classification into the ``at-risk'' group. The number of possible binary trees is much too large for brute force enumeration\footnote{In our application our item bank consists of 173 items, each with up to 6 possible cutpoints.}; many possible heuristics are available, and different procedures will lead to higher-utility screening instruments than others. Here we focus on one method for populating our action space, motivated by the Bayesian decision theory context. 

We first obtain a regression tree $T$ by applying a particular tree growing algorithm (described shortly) to large Monte Carlo samples from the posterior predictive distribution $f(\tilde{\mathrm{x}},\tilde{y}\mid \mathrm{x}_{1:n}, y_{1:n})$. We then choose the cutoff $C_T$ that optimizes the expected utility (\ref{eqn:expected_utility_intuitive}) relative to $T$ over these samples.

Our proposed heuristic for obtaining the regression tree $T$ relies on a novel stopping criterion we call \emph{maxIPP}, for ``\textbf{max}imum \textbf{I}tems \textbf{P}er \textbf{P}ath.'' The maxIPP criterion denotes the maximum number of \emph{unique} items in each root-to-leaf path of the decision tree defining the adaptive test, and consequently, the number of items each individual will be administered during their screening test. The tree is grown using a variation of the CART algorithm; it achieves the maxIPP constraint by restricting the items available for splitting in a given path after $m$ unique items have been used. For details, see Appendix \ref{Appendix:maxIPP criterion}. 

Categorizing trees by maxIPP is useful in our context of shortening lengthy instruments. While maximum depth also limits the number of items, maxIPP allows for further splitting on items already administered, without counting them against the tree ``cost''. 

For a given $m$, we calibrate an approximately optimal tree with maxIPP $m$ (denoted $T^*_m$) to synthetic data drawn from the posterior of $\theta_\mathrm{X}$ and the posterior predictive of $\tilde{\mathrm{X}}$. Specifically, our synthetic data used for calibrating $T^*_m$ are $\{\tilde{\mathrm{x}}_{k}, \bar{\mathbb{E}}(\tilde{Y} \mid \tilde{\mathrm{x}}_k)\}$, $1\leq k \leq M$, where the second element is the posterior predictive ``at-risk'' probability, given $\tilde{\mathrm{x}}_k$:
\begin{align}
\label{eqn:postmean_formal}
\bar{\mathbb{E}}(\tilde{Y} \mid \tilde{\mathrm{x}}_k) = \int_{\Theta_Y} \mathbb{E}(\tilde{Y} \mid \tilde{\mathrm{x}}_k, \theta_Y) \pi(\theta_Y \mid \mathrm{x}_{1:n},y_{1:n}) \; d\theta_Y.
\end{align}
As a reminder, $\pi(\theta \mid \mathrm{x}_{1:n}, y_{1:n})$ is the posterior density of $\theta$, having observed data $(\mathrm{x}_{1:n}, y_{1:n})$. We use the term ``calibrate'' rather than ``fit'' for the process of applying the maxIPP algorithm to synthetic data, in order to reserve the term ``fit'' for the context of fitting the Bayesian models to real data. 

Having obtained $T^*_m$, the cutoff $C_{T^*_m}$ is then optimized relative to the (unconditional) posterior predictive expected utility:
\begin{align*}
C_{T^*_m}&=\argmax_{C\in[0,1]}\mathbb{E}U(\text{Thr}_C(T^*_m))\\
&=\argmax_{C\in[0,1]} [w\cdot\text{Sensitivity}(\text{Thr}_C(T^*_m)) + (1-w)\cdot\text{Sensitivity}(\text{Thr}_C(T^*_m))],
\end{align*}
where the inner expression on the right-hand side (i.e., the weighted average of sensitivity and specificity of $\text{Thr}_{C}(T^*_m)$) is approximated using
\begin{equation}
\small
\label{eqn:computing_utility_from_sample_all}
 w\cdot \frac{\sum_{k=1}^M\mathbbm{1}(\text{Thr}_C(T^*_m(\tilde{\mathrm{x}}_{k}))=1, \tilde{y}_{k}=1)}{\sum_{k=1}^M\mathbbm{1}( \tilde{y}_{k}=1)} + (1-w)\cdot \frac{\sum_{k=1}^M\mathbbm{1}(\text{Thr}_C(T(\tilde{\mathrm{x}}_{k}))=0, \tilde{y}_{k}=0)}{\sum_{k=1}^M\mathbbm{1}( \tilde{y}_{k}=0)}.
\end{equation}

In summary, $T_m^*$ is our final regression tree with maxIPP $m$ that predicts the probability of being ``at-risk" given a set of item responses, and $\text{Thr}_{C_{T_m^*}}$ maps these probabilities to a predicted class status 0 or 1 ($0=$ ``at-risk'', $1=$ ``not-at-risk''). The threshold is chosen relative to the specific regression tree $T_m^*$, to optimize the utility function for the target population. We use $\gamma^*_m=\text{Thr}_{C_{T_m^*}}(T_m^*)$ to denote our approximately optimal tree-based adaptive test of length $m$.  

Our action space $\Gamma$ consists of one adaptive test $\gamma^*_m$ for each value of $m$ under consideration for a given application. We emphasize this is just one proposed heuristic for obtaining an adaptive screening test that optimizes Equation (\ref{eqn:expected_utility_intuitive}), while administering at most $m$ items. One can obtain adaptive tests with $m$ items using other tree growing methods calibrated with other synthetic or real data. Each of these can be compared using the criteria described in Section \ref{section:Comparing adaptive tests} before choosing a final adaptive test; see Section 4 of the Supplementary Material (\cite{Krantsevich2022}) for comparisons of several methods.


\subsubsection{Comparing adaptive tests of different lengths}
\label{section:Comparing adaptive tests}
Once we have (at least) one action $\gamma^*_m$ for each test length $m$, we need to choose the value of $m$ for the final adaptive screening test. In general, shorter screening tests can only degrade accuracy (utility), so the relevant questions are ``by how much?'' and ``with what statistical uncertainty''? 

To address these questions we define a random variable (with respect to the posterior distribution) $\Delta_{\theta,m}$ that characterizes the utility loss due to shortening to $m$ questions.
That is, we are interested in the difference in expected utility between that of the shortened exam $\mathbb{E}U_\theta(\gamma^*_m)$ and that of the full, non-shortened, exam $\mathbb{E}U_\theta(\gamma^{*})$. Here, the optimal non-shortened action is $\gamma^*(\cdot) = \text{Thr}_{C^*}(\bar{\mathbb{E}}(\tilde{Y} \mid \cdot))$, where $\bar{\mathbb{E}}(\tilde{Y} \mid \cdot)$ is as in (\ref{eqn:postmean_formal}), and $\text{Thr}_{C^*}$ is optimized relative to the posterior predictive expected utility; specifically,  $\text{Thr}_{C^*}$ is optimized using Equation (\ref{eqn:computing_utility_from_sample_all}), but with $\bar{\mathbb{E}}(\tilde{Y} \mid \tilde{\mathrm{x}})$ in place of $T^*_m(\tilde{\mathrm{x}})$. We denote this difference as 
\begin{equation}
\Delta_{\theta,m} = \mathbb{E}U_\theta(\gamma^*_m) - \mathbb{E}U_\theta(\gamma^*).
\end{equation}
To obtain Monte Carlo samples of $\Delta_{\theta, m}$, for each posterior draw $\theta^{(j)}$ compute
\begin{equation*}
\Delta_{\theta^{(j)},m}=\mathbb{E}U_{\theta^{(j)}}(\gamma^*_m) - \mathbb{E}U_{\theta^{(j)}}(\gamma^*),
\end{equation*}
where $\mathbb{E}U_{\theta^{(j)}}(\gamma)$ is computed using (\ref{eqn:computing_utility_from_sample}). 

Boxplots may then be plotted for each value of $m$. See Figure \ref{fig:changing_utility_function} for an example with boxplots of $\Delta_{\theta, m}$ varying the number of items $m$ and the weight $w$ that defines the utility function $U$.These utility difference plots visually represent our statistical uncertainty of the trade-offs between assessment sensitivity/specificity and length.

\subsection{Comparison to existing methods for designing adaptive tests}

In Section \ref{section:Populating_the_action_space} we proposed a novel algorithm for obtaining adaptive tests of different lengths to populate the action space, our second main contribution. Here we compare to existing work on tree-based adaptive tests, as an IRT-based test is not appropriate for our application (see Section \ref{section:CART}). For comparisons between tree-based adaptive tests and IRT, see \cite{Gibbons2016} and \cite{Zheng2020}. 

To the best of our knowledge, current tree-based adaptive tests are fit using existing algorithms like CART; built-in hyperparameters decide the length of the test, and the optimization criteria (typically Gini index) is not specific to the adaptive testing context. Typically, the decision tree is fit to real item response - outcome data. Two exceptions are \cite{Gibbons2016}, who fit the tree to locally perturbed artificial data for increased model stability, and \cite{Zheng2020}, who utilized SMOTE to help with class imbalance.

The purpose of synthetic data in our application is to provide an MCMC approximation of the expected utility integral over the target population. We obtain this data by modeling the high-dimensional joint density of item responses -- risk outcome via two sophisticated Bayesian models, and use a context-specific utility function (i.e. sensitivity and specificity) for tree optimization, rather than Gini index. Finally, our novel maxIPP stopping criterion is an application-specific design choice, exploiting the fact that items can be reused for splitting.


\section{Screening for Youth delinquency in Honduras}
\label{section:Youth delinquency in Honduras}
\subsection{Data}
\label{section:Data}

The instrument used to collect data for this project was the Instrumento de Medicion de Comportamientos (IMC), a revised version of the original Honduran YSET, which was itself a Spanish adaptation of the YSET developed by \cite{Hennigan2014}. Under a collaboration with the Center for Violence Prevention and Community Safety at Arizona State University, the item bank for the Honduran YSET was expanded to include protective factors and increase the number of risk factors measured, drawing on the Communities That Care survey, Eurogang Youth Survey, and others. This revised item bank was further refined to increase predictive power in the local context.

Our data consists of responses to the IMC from 3972 school-attending youth. The IMC covers basic demographics about the youth, along with 173 items measuring 38 risk and protective factors over four domains: community, family, school and peer/individual. The risk and protective factor scales are provided in Table \ref{table:IMC_factors}. Our variable $\mathrm{X}$ consists of responses to these 173 items. 

Our data also include answers to 18 items that measure seven problem behaviors. Three items measure \emph{violent behavior}, four items measure \emph{property crime}, three items measure \emph{gang involvement}, three items measure \emph{alcohol and drug use}, two items measure \emph{drug sales}, two items measure \emph{weapons carrying}, and one item measures \emph{truancy}.  In what follows, the outcome $Y$ is a binary variable denoting whether or not the youth is at risk of \emph{violent behavior}. The three items related to \emph{violent behavior} are:
\begin{quote}
\normalsize
\begin{enumerate}
	\item In the past 6 months, have you hit someone with the intention of hurting them?
	\item In the past 6 months, have you attacked someone with a weapon?
	\item In the past 6 months, have you used a weapon or force to get money or goods from someone? 
\end{enumerate}
\end{quote}
Youth are deemed to be ``at-risk'' $(Y=1)$ if they answer ``yes'' to any of the three items above. Items measuring the other six problem behaviors are not utilized for this analysis.
\newline \newline
\noindent \emph{Connection to previous notation.} We have responses to the 173 items $\mathrm{X}$ and an outcome variable $Y$ denoting whether or not the youth is in the ``at-risk'' group for violent behavior  (``at-risk'' = 1,``not-at-risk'' = 0). The variable $\gamma$ denotes an adaptive test which takes the youth's responses to a subset of the 173 items and predicts a risk class. For our purposes, $\gamma$ is composed of two parts: a binary decision tree $T$ that maps item responses to a risk probability, and a threshold $C$ that determines risk class based on risk probability. We will analyze the quality of a risk assessment $\gamma$ using an expected utility function $\mathbb{E}U$; $\mathbb{E}U(\gamma)$  is a weighted average of the sensitivity and specificity of the risk assessment $\gamma$.

\subsection{Modeling}
\label{section:Modeling}
We model the data $(\mathrm{X}, Y)$ compositionally as $f(\mathrm{x}, y) = f(y \mid \mathrm{x})f(\mathrm{x})$. We model $f(\mathrm{x})$ as a Gaussian copula factor model and $f(y \mid \mathrm{x})$ as a logistic XBART model using the \texttt{bfa} and  \texttt{xbart} packages, respectively; this model specification is quite flexible. See Appendix \ref{Appendix:Integrating} for details on integrating over $f(\tilde{\mathrm{x}}, \tilde{y}\mid \theta^{(j)})$ using the fitted models. 

As with any Bayesian modeling endeavor, we recommend interrogating model quality and adjusting hyperparameters accordingly via standard posterior predictive checks, including plots to avoid model misspecification; see, for example, \cite{Gelman1996} and \cite{Gabry2019}. These model checks should be performed on training data, and not adjusted after obtaining results on hold-out or validation data. This is the approach we used to determine the number of factors in the model for $f(x)$.

\subsubsection{Item responses}

The model we use for $f(\mathrm{x})$ is a Gaussian copula factor model (GCFM), proposed by \cite{Murray2013} and implemented in the R package \texttt{bfa}. Gaussian copula factor models unite Gaussian factor models with the Gaussian copula. The joint distribution of the fitted model assumes the dependence structure of the Gaussian factor model, but with marginal distributions estimated nonparametrically from the data. The joint dependence structure of the Gaussian factor model is reasonable considering the factor-based nature of the latent constructs being measured by adaptive tests. Additionally, the nonparametric estimation of the marginal distributions is an advantage over methods that assume normal marginals. See Section 3.1 of the Supplementary Material (\cite{Krantsevich2022}) for details.

Note that the GCFM was fit to an augmented vector including age: $(\mathrm{X}, \text{Age})$. This allows us to condition on age in defining the posterior prediction distribution that represents our target population. While we could have accomplished this by only fitting the GCFM to data from a particular age group, fitting the model to the entire population and sampling conditionally after the fact allows for borrowing information from the larger population, and deploying it in service of a subpopulation with fewer data. We only use item responses as splitting variables (inputs) for the adaptive tests.

Sensitivity analysis to the number of factors via posterior predictive checks revealed that 3 or more factors yielded similar conclusions; results for the $k=3$ factor specification in the Gaussian factor copula model are reported here.

\subsubsection{Risk prediction}

We model $f(y\mid \mathrm{x})$ using a log-linear Accelerated Bayesian Additive Regression Trees (XBART) model that builds on the log-linear Bayesian Additive Regression Trees (BART) model for multinomial logistic regression of \cite{Murray2020} with a modification of the ``accelerated'' model fitting algorithm of \cite{He2018} developed by \cite{Wang2021}. See Section 3.2 of the Supplementary Material (\cite{Krantsevich2022}) for details. The log-linear XBART classification model provides class probability predictions $\bar{\mathbb{E}}(\tilde{Y} \mid \tilde{\mathrm{x}}_k)$, the probability that a youth with item responses $\tilde{\mathrm{x}}_k$ is in the ``at-risk'' group. This modeling choice provides the predictive accuracy and Bayesian uncertainty quantification abilities of BART-based models, with the computational speed-up of the XBART family and the classification-specific adaptions implemented by \cite{Wang2021}. Notably, this approach is substantially less constrained than typical IRT approaches, which require that the risk probability relates to the item response via the same low-dimensional latent factors. Here, while we assume that the item responses have a latent dimension of $k = 3$, the risk probability can depend directly on every single item individually (with no dimension reduction). However, regularization priors in the tree ensemble representation favor trees that utilize far fewer than every available item.

\subsubsection{Connection to previous notation} 
We fit the Gaussian copula factor model to item responses and age from the IMC data, then obtain synthetic item response data $\{\mathrm{\tilde{x}}_k\}_{k=1}^M$ from the target population using the predictive distribution of the fitted model.
 
 The plots in the next section compare the utility of the shortened screening test $\gamma_m^*$ to the utility of the full-length test $\gamma^*$, which uses all 173 items on the IMC. The instrument $\gamma^*(\cdot) = \text{Thr}_{C^*}(\bar{\mathbb{E}}(\tilde{Y} \mid \cdot))$ is composed of a regression function $\bar{\mathbb{E}}(\tilde{Y} \mid \cdot)$ predicting the probability of the Honduran youth being ``at-risk'', followed by a thresholding function $\text{Thr}_{C^*}$ to predict risk class status. We fit the regression function as an XBART model using the IMC data, then obtain predicted ``at-risk'' probabilities $\bar{\mathbb{E}}(\tilde{Y}\mid \tilde{\mathrm{x}}_k)$ for the synthetic item responses $\mathrm{\tilde{x}_k}$. The thresholding function is chosen to optimize the utility function for the target population, given predicted ``at-risk'' probability $\bar{\mathbb{E}}(\tilde{Y}\mid \tilde{\mathrm{x}}_k)$.

The shortened instrument with $m$ items, $\gamma^*_m(\cdot) = \text{Thr}_{C_{T^*_m}}(T_m^*(\cdot))$, is composed of a binary regression tree $T_m^*$ with maxIPP $=m$, and a thresholding function $\text{Thr}_{C_{T^*_m}}$. The adapted test $T^*_m$ is calibrated using synthetic data $\{\tilde{\mathrm{x}}_{k}, \bar{\mathbb{E}}(\tilde{Y} \mid \tilde{\mathrm{x}}_k)\}$, $1\leq k \leq M$. The thresholding function for $\gamma_m^*$ is computed similarly to the one for $\gamma^*$, except that it optimizes the cutoff using ``at-risk'' probabilities $T^*_m(\tilde{\mathrm{x}}_k)$ rather than $\bar{\mathbb{E}}(\tilde{Y}\mid \tilde{\mathrm{x}}_k)$. 

\section{Results}
\label{section:Results}
Section \ref{section:Demonstration_of_the_method} provides a demonstration of the method using the data for youth delinquency in Honduras. Section \ref{section:Out of sample validation} provides out-of-sample validation of the method on a hold out set, along with a subgroup analysis using the same hold-out set. 

\subsection{Demonstration of the method}
\label{section:Demonstration_of_the_method}


Recall the three steps in the decision theory framework laid out above: 1) a utility function for measuring the ``goodness'' of the assessment; 2) a target population; 3) a method for obtaining assessments of different lengths. Sections \ref{section:changing utility}, \ref{section:changing target population}, and \ref{section:changing action space} demonstrate how the utility difference plots change as we vary these three choices, respectively, when applied to the Honduras youth risk assessment data.

\subsubsection{Changing the utility function} 
\label{section:changing utility}
\begin{figure}
    \includegraphics[width=0.8\linewidth]{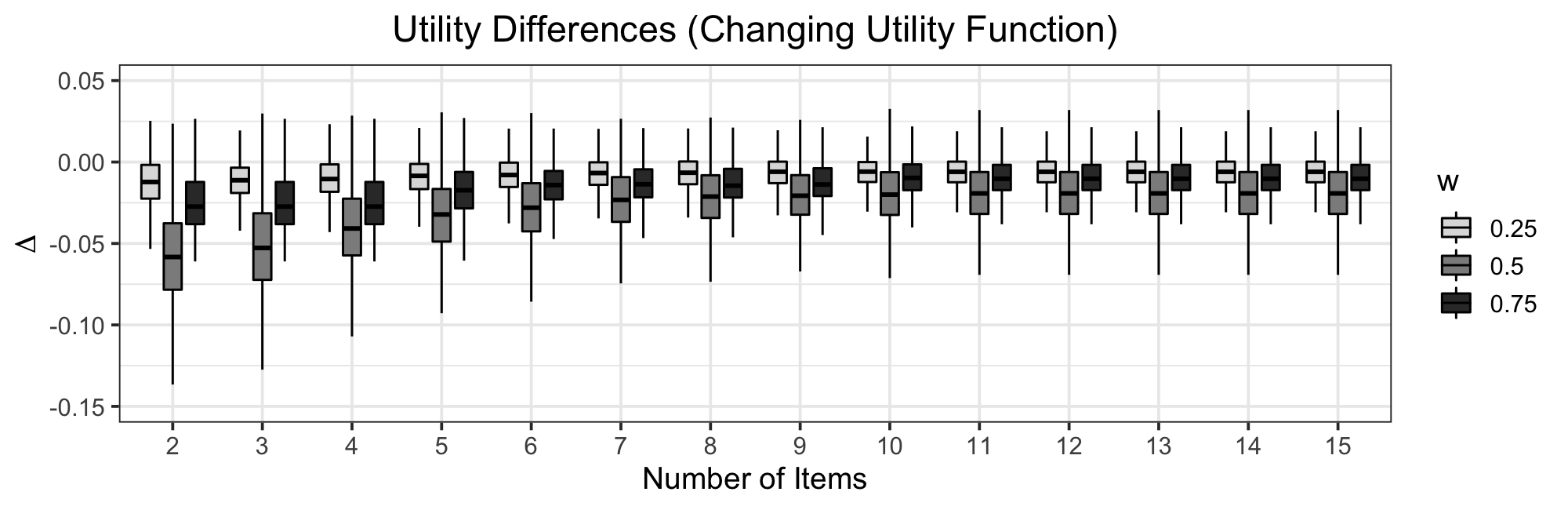} 
    \caption{Each boxplot represents the posterior distribution of $\Delta_{\theta,m}=\mathbb{E}U_{\theta}(\gamma^*_m) - \mathbb{E}U_{\theta}(\gamma^*)$ for a particular number of items $m$, and a particular value of $w$ in the utility function from Equation \ref{eqn:expected_utility_intuitive}. The samples of $\Delta_{\theta,m}$ that form the boxplot are obtained by computing $\mathbb{E}U_{\theta^{(j)}}(\gamma)$ (via sensitivity and specificity) for $\gamma_m^*$ and $\gamma^*$, on a synthetic data sample $\{\tilde{\mathrm{x}}_{ij}, \tilde{y}_{ij} \mid \theta^{(j)}\}_{i=1}^N$. This sample is obtained from posterior draw $\theta^{(j)} = (\theta_\mathrm{X}^{(j)}, \theta_Y^{(j)})$ of the two Bayesian models. }
    \label{fig:changing_utility_function}
\end{figure}

First, we highlight how the plots change when we vary Step 1, the utility function. Figure \ref{fig:changing_utility_function} shows boxplots of the difference in expected utility for three different weights $w$ in the utility function 
\begin{equation*}
\mathbb{E}U_\theta(\gamma)=w\cdot \text{Sensitivity}(\gamma) + (1-w)\cdot \text{Specificity}(\gamma).
\end{equation*} 

For each weight $w$ and each value of $m$, we compute draws of the utility difference $\Delta_{\theta^{(j)}, m}=\mathbb{E}U_{\theta^{(j)}}(\gamma^*_m) - \mathbb{E}U_{\theta^{(j)}}(\gamma^*)$ using synthetic data from each posterior draw $j$; the posterior distribution of $\Delta_{\theta, m}$ is then visualized via a boxplot of the draws $\{\Delta_{\theta^{(j)}, m}\}_{j=1}^D$. The distribution of $\Delta_{\theta,m}$ can vary depending on our choice of both $m$ and $w$. 

\begin{figure}
    \includegraphics[width=\linewidth]{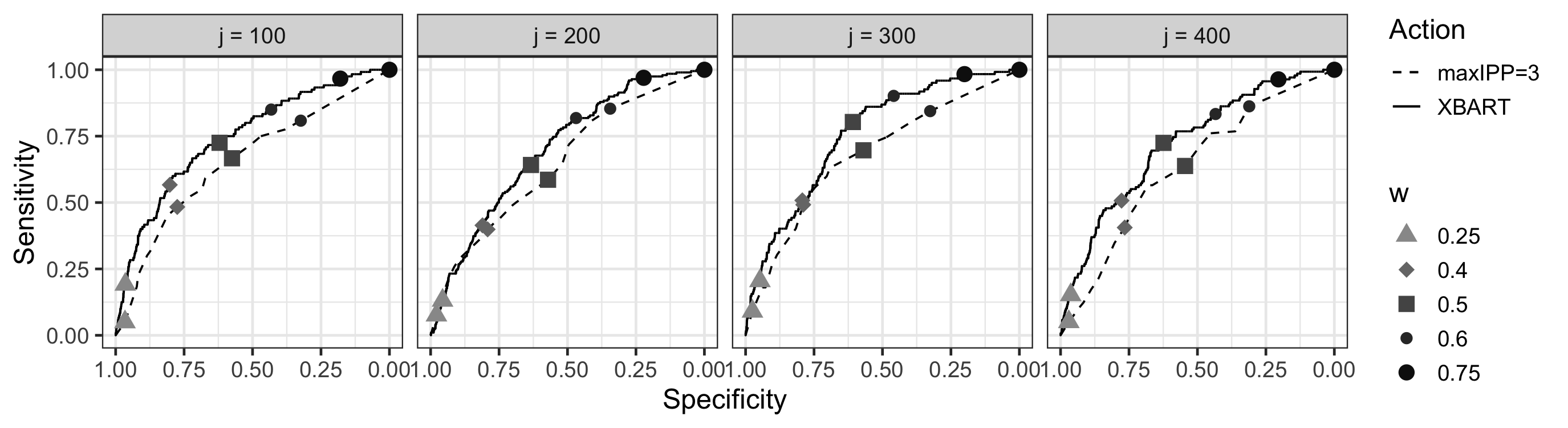} 
     \caption{Here we show how to obtain posterior draws of $\Delta_{\theta^{(j)}, m}$ from posterior draws $\theta^{(j)}$. The four plots show ROC curves obtained from synthetic data samples $\{\tilde{\mathrm{x}}_{ij}, \tilde{y}_{ij} \mid \theta^{(j)}\}_{i=1}^N$ for $j=100, 200, 300, 400$. The values of $j$ shown here were arbitrarily chosen for demonstration and are not inherently important. The ROC curves are computed based on the predicted ``at-risk'' probabilities $\bar{\mathbb{E}}(\tilde{Y}\mid \tilde{\mathrm{x}}_{ij})$ and $T^*_m(\tilde{\mathrm{x}}_{ij})$ from the XBART action $\gamma^*$ and  maxIPP $=3$ action $\gamma_{m=3}^*$ (respectively) for each specific $j^{th}$ population. For each given $w$, there is exactly one cutoff $C$ which maximizes the utility function $\mathbb{E}U_\theta(\gamma)=w\cdot \text{Sensitivity}(\gamma) + (1-w)\cdot \text{Specificity}(\gamma)$ over all sample populations (all values of $j$), for $\gamma = \gamma_{m=3}^* = \text{Thr}_{C_{T^*_m}}(T_m^*(\cdot))$. That cutoff $C$ corresponds to a particular (Specificity, Sensitivity) pair for each value of $j$ (for both the XBART and maxIPP$=3$ actions), which are visualized as points on the ROC curves from those two actions for the $j^{th}$ synthetic population. Those Sensitivities and Specificities are used to compute the realized utility values $\mathbb{E}U_{\theta^{(j)}}(\gamma^*_m)$ and $\mathbb{E}U_{\theta^{(j)}}(\gamma^*)$, along with their difference,  $\Delta_{\theta^{(j)},m}$, which contributes one point to the boxplots in Figure \ref{fig:changing_utility_function} for the given values of $w$ and $m=3$. Notice that values of $w$ closer to 1 lead to differences in sensitivity between $\gamma^*$ and $\gamma_{m=3}^*$ (distance between points on the Sensitivity axis) being smaller than differences in specificity (distance between points on the Specificity axis). This can be observed in the points corresponding to $w=0.75$ and, to a lesser extent, $w=0.6$. The opposite behavior is observed for $w=0.25$ and $w=0.4$.}
    \label{fig:ROC_with_w}
\end{figure}

Figure \ref{fig:ROC_with_w} provides a visual example of how (Specificity$(\gamma)$, Sensitivity$(\gamma)$) for $\gamma\in \{\gamma^*, \gamma^*_m\}$, in conjunction with $w$, lead to different draws of $\Delta_{\theta,m}$ for $m=3$. In particular, as $w$ gets closer to 0 or 1, it is easier for the shortened test $\gamma^*_m$ to achieve a utility value closer to that of the non-shortened instrument $\gamma^*$. Practically, a value of $w$ close to 0 or 1 amounts to strongly favoring either sensitivity or specificity, at the expense of the other; such decisions can have unintended ramifications, which are discussed further in Section \ref{section:disparate_impact}.

\subsubsection{Changing the target population}
\label{section:changing target population}
\begin{figure}
	\includegraphics[width=0.8\linewidth]{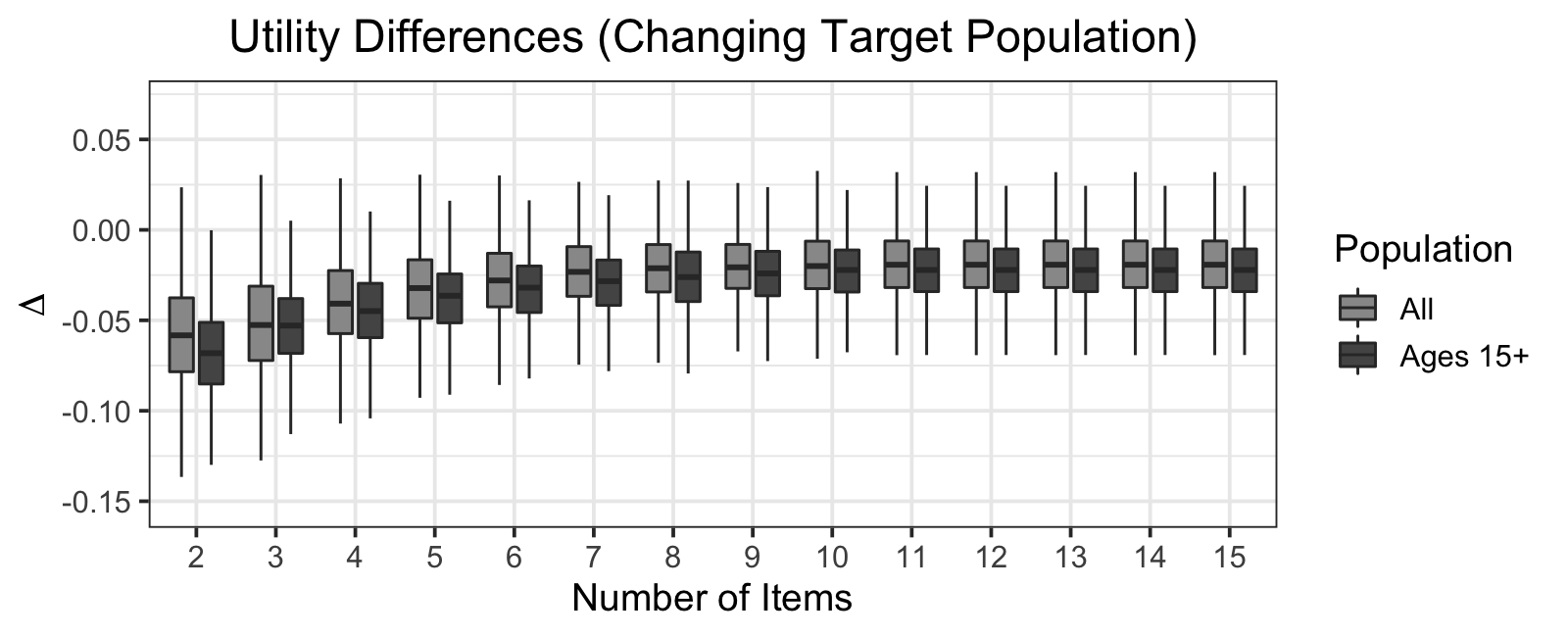}
	\caption{Utility difference plots when calibrating the trees to a different target population. Plots here are shown for a target population being every youth in the full IMC data (``All''), and those youth ages 15 and older (``Ages 15+''). The utility plots are quite similar, although exam truncation seems to result in a greater loss of utility (relative to the full screening instrument) for the older group. Calibrating the adaptive test to the subgroup also results in slightly more certainty compared to the full population.}
	\label{fig:Utility_diffs_population}
\end{figure}

Next, we vary Step 2, the target population. The boxplots in Figure \ref{fig:Utility_diffs_population} represent the same quantity as Figure \ref{fig:changing_utility_function} (namely, the distribution of $\Delta_{\theta,m}$). However, Figure \ref{fig:Utility_diffs_population} shows expected utility differences for adaptive tests calibrated using two target populations: all Honduran youth, and youth ages 15 and older. We chose to target youth ages 15 and older since age 15 marks the transition from middle school to secondary school, as well as the $\text{quincea}\tilde{\text{n}}\text{era}$ ceremony. To change the target population, we used the GCFM fit to the entire dataset, but then drew samples $\{\tilde{\mathrm{x}}_{ij}, \tilde{y}_{ij}\mid \theta^{(j)}\}$ using the conditional predictive distribution, $f(\tilde{\mathrm{x}}, \tilde{y}\mid \mathrm{x}_{1:n}, y_{1:n}, \text{Age}\geq15)$. 

\begin{figure}
    \includegraphics[width=0.7\linewidth]{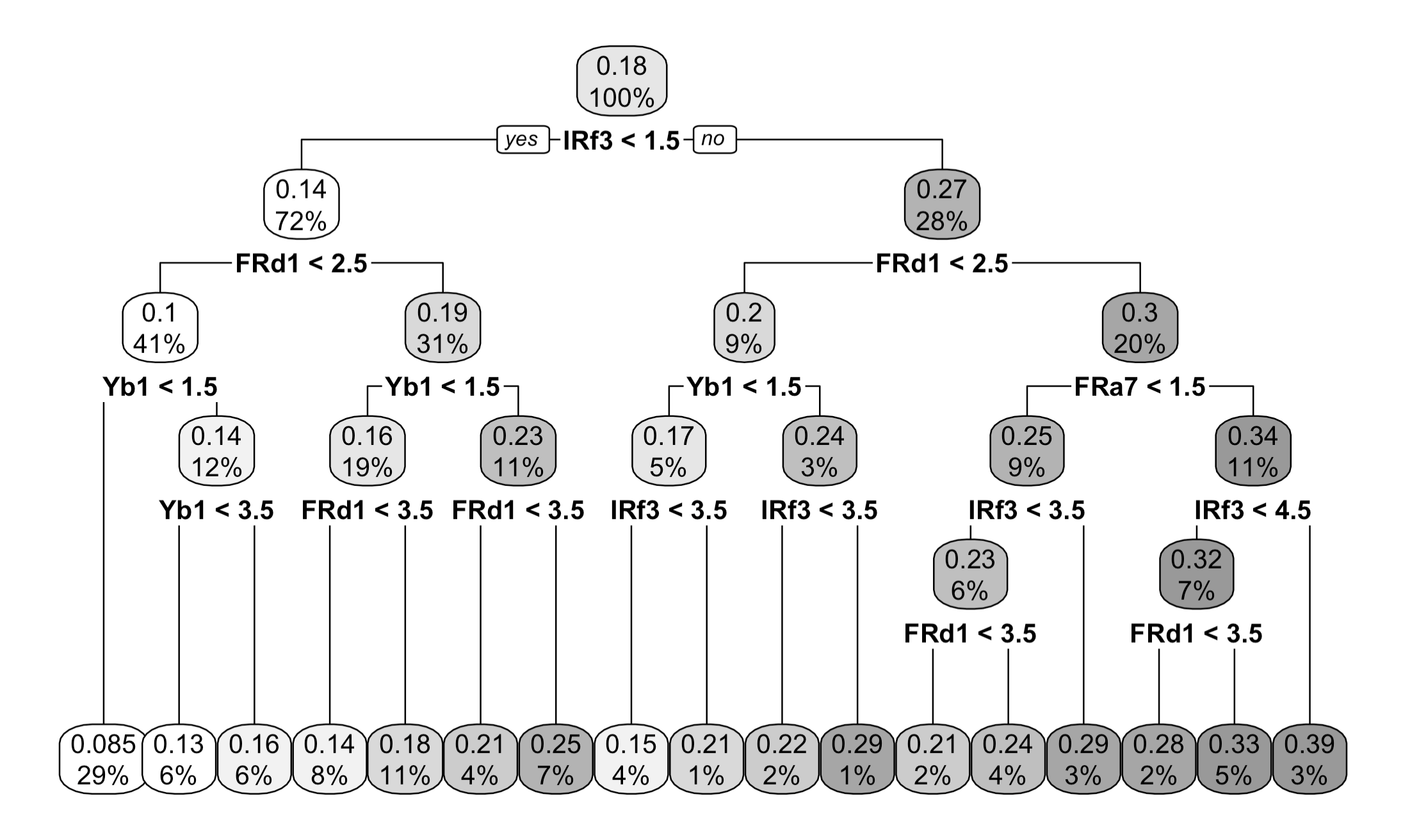} 
    \caption{Tree representing the adaptive test calibrated using the entire group of Honduran youth as the target population. The items and item responses corresponding to each node label and cutpoint, respectively, are found in Table \ref{table:Items}. This figure and Figure \ref{fig:Tree_subgroup} were created using the \texttt{rpart.plot} package (\cite{Milborrow2021}).}
    \label{fig:Tree_all}
\end{figure}

\begin{figure}
    \includegraphics[width=0.7\linewidth]{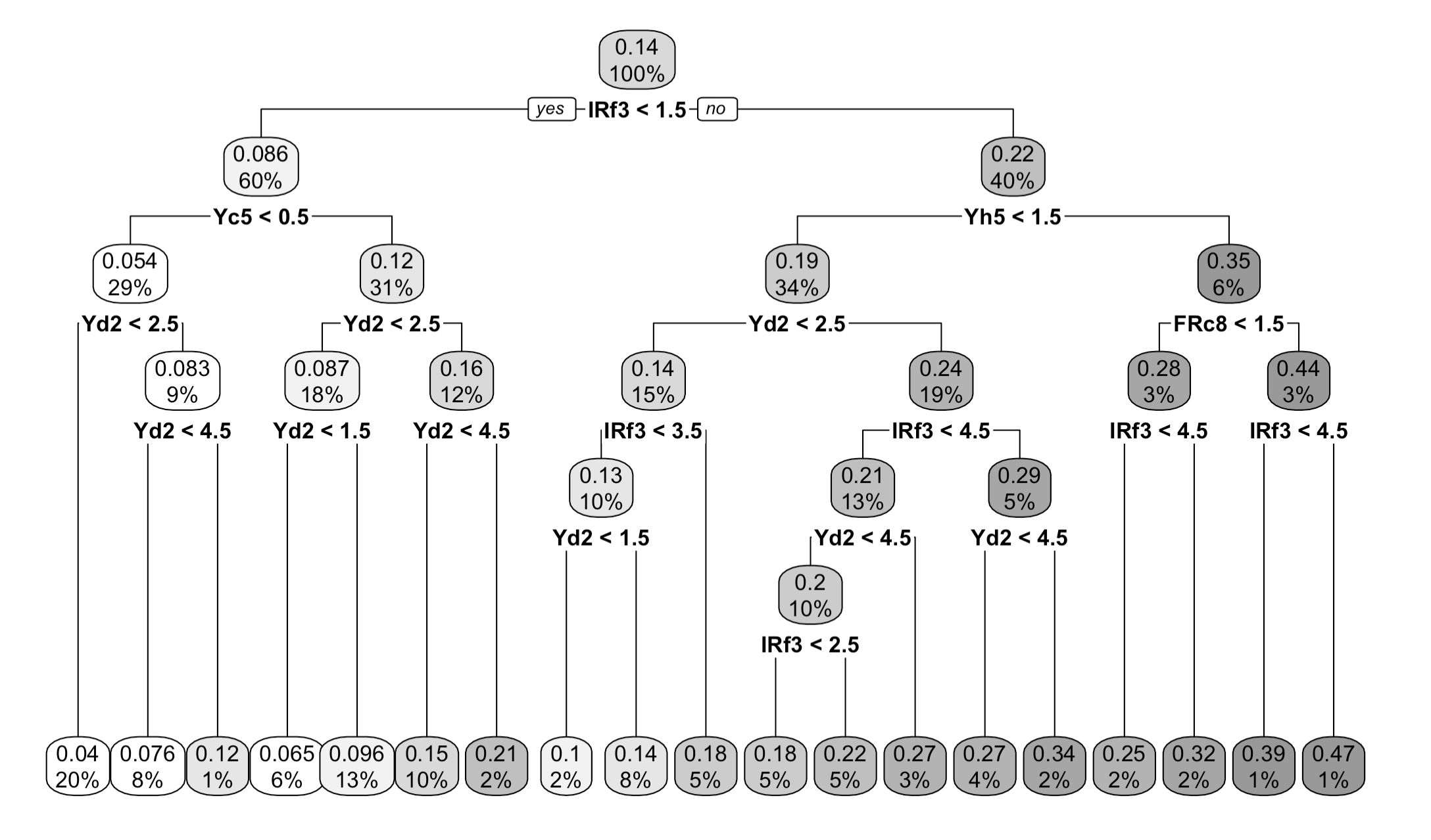} 
    \caption{Tree representing the adaptive test calibrated using the group of Honduran youth ages 15 and older as the target population. The items and responses corresponding to the node labels and cutpoints are found in Table \ref{table:Items}. }
    \label{fig:Tree_subgroup}
\end{figure}

The expected utility plots are similar; however, targeting the subgroup when designing the adaptive test yields slightly less variability in the posterior estimates of the utility difference. Interestingly, these similar results arise based on adaptive tests that use different splitting items and cutpoints. Figures \ref{fig:Tree_all} and \ref{fig:Tree_subgroup} show the trees with maxIPP of $3$ representing the adaptive tests for these two target populations. The items corresponding to these trees are listed in Table \ref{table:Items}, with the response options found in Table \ref{table:Items_appendix}. Notice that because of the maxIPP criterion, these trees have a maximum depth of 5, but have only 3 unique items in each root-to-leaf path.

\begin{table}[h!]
\centering
\caption{Items corresponding to the splitting variables present in the tree-based adaptive tests of Figures \ref{fig:Tree_all} and \ref{fig:Tree_subgroup}. The right-most column shows the target population for which this variable is included in its adaptive test.}
\label{table:Items}
\begin{tabular}{c l l c} 
 \hline 
 \textbf{Variable} & \textbf{Item} & \textbf{Response Options } & \textbf{Population(s)} \\  
 \hline
IRf3 &  In the past 6 months, how many of your best  & 1 = None of my friends & All Youth \\ 
& friends have tried beer, wine or hard liquor & 2 = 1 of my friends & \\
& (for example, vodka, whiskey or gin) when &  3 = 2 of my friends & \\
& their parents didn’t know  about it? &  4 = 3 of my friends & \\
& & 5 = 4 of my friends or more & \\
\hline
Yc5 &  In the last year, have you fought or had a  & 0 = No & All Youth \\ 
& problem with a friend?  & 1 = Yes & \\
\hline
FRd3\_ctc &  People in my family often insult or yell at   & 1 = No! & All Youth\\
& each other. & 2 = no  & \\ 
&& 3 = yes & \\
&& 4 = Yes!  & \\
\hline
FRa6 & Has anyone in your family had a severe alcohol & 1 = No  & All Youth   \\ 
&or drug problem?  & 2 = Yes & Age $\geq 15$\\
\hline
Yh5 &  During the last six months, how many friends   & 1 = None  & All Youth \\ 
 & have belonged to or have joined a gang or & 2 = A few  & Age $\geq 15$ \\
 & ``mara''?  &  3 = Half & \\
 && 4 = Most & \\
 && 5 = All  & \\ 
\hline
Yd2 & Sometimes I find it exciting to do things that & 1 = Strongly disagree & Age $\geq 15$  \\  
&  could get me in trouble.  & 2 = Disagree    & \\
&&  3 = Neither agree or disagree   & \\
&&   4 = Agree  & \\
&&   5 = Strongly agree  & \\

\hline
Ya6 &  People “blame me” for lying or cheating. & 1 = Never & Age $\geq 15$ \\ 
&&  2 = Rarely   & \\
&&  3 = Half the time   & \\
&&    4 = Often  & \\
&& 5 = Always    & \\
 \hline
\end{tabular}
\end{table}

\subsubsection{Changing the algorithm to populate the action space}
\label{section:changing action space}
Finally, we can consider different algorithms for populating the action space. In this paper we have focused on the composite action $\gamma = \text{Thr}_C(T)$, a regression tree $T$ predicting ``at-risk'' probability followed by a cutoff $C$ that determines risk status. Our proposed method for populating the action space is a regression tree obtained by applying the maxIPP growing and pruning method to synthetic data obtained from the posterior predictive distribution, and a threshold optimized to the utility function for the tree $T$.

Many other methods are possible. For example, one can calibrate the regression tree using a stopping criterion like maximum depth, or apply the algorithm to different synthetic data or to real data; a classification tree can be used as the adaptive test directly instead of a regression tree followed by a cutoff. We explore these possibilities in Section 4 of the Supplementary Material (\cite{Krantsevich2022}). The main takeaway is that tree-based adaptive tests that do not optimize the utility function at all during their design are significantly worse at reproducing the utility of a full-item assessment, relative to adaptive tests that do. 

\subsection{Out-of-sample corroboration}
\label{section:Out of sample validation}
The proposed method will be empirically reliable only insofar as the posterior predictive distribution suitably reflects the distribution of future outcomes. To verify that our Gaussian copula factor and XBART models are succeeding in this regard, we perform the following hold-out experiment. Our data was collected in two different time periods, the first wave between September and November of 2017 and the second wave between January and February of 2018. 
The earlier-collected data is our training set and consists of 2787 youth; the later data is our testing set and consists of 1185 youth. Simply put, this experiment answers the question: how would our approach have performed if we had applied it in 2018, based on the 2017 data?

\begin{figure}
	\includegraphics[width=0.8\linewidth]{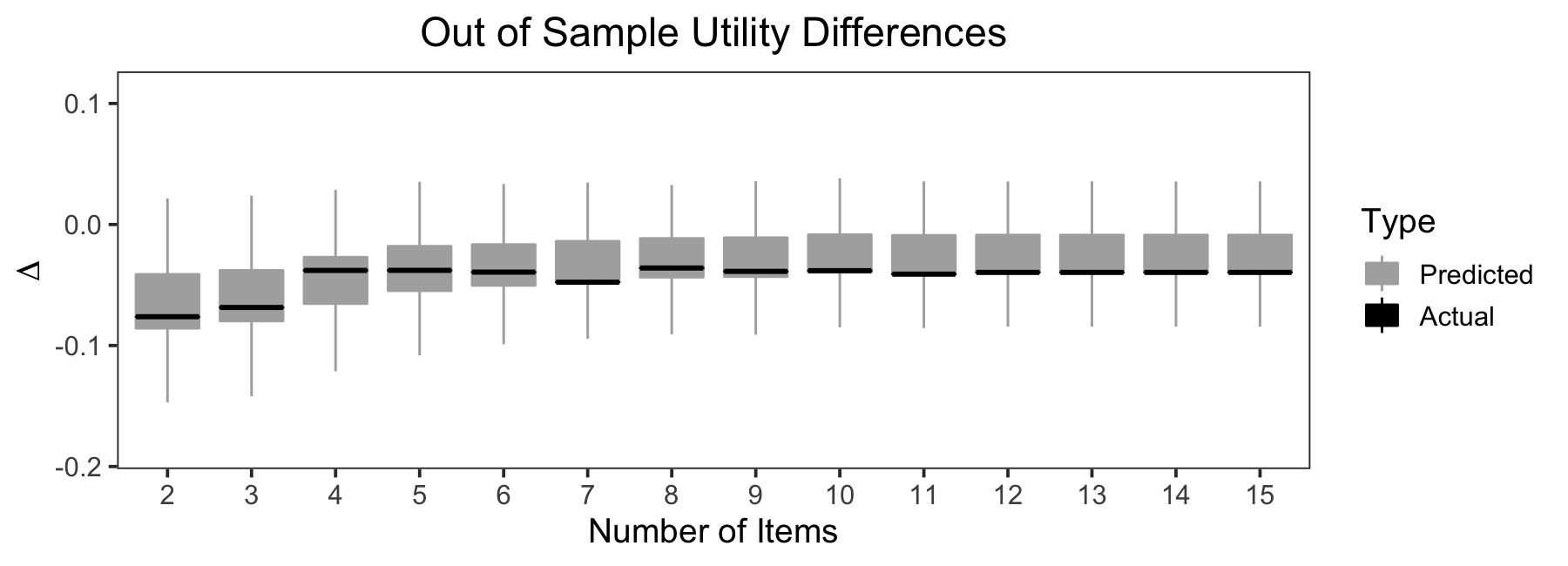}
	\caption{Plots of the projected difference in expected utility produced via our method on the training data, and the actual difference in expected utility computed on the test data; these are results for $w=0.5$.}
	\label{fig:Utility_diffs_train_test}
\end{figure}

Figure \ref{fig:Utility_diffs_train_test} demonstrates the expected utility difference plots we obtained by applying our method on the training data, and the actual expected utility differences on the testing data. To compute the actual expected utility, we used our proposed method on the training data to obtain a tree-based adaptive test for each value of maxIPP, along with a full-item (non-shortened) test. We also produced the boxplots representing our uncertainty around $\Delta_{\theta, m}$ using the training data alone. We then predicted risk classes on the testing set using both the tree-based adaptive test and the full item test, and computed the difference in empirical utility over the testing set. The empirical utility on the testing set is always within our predicted range, in fact within the $25^{th}$ and $75^{th}$ quantiles of the distribution. 

Beyond utility differences relative to the full item test, practitioners are interested in the absolute sensitivity and specificity of the instrument. Table \ref{table:oos_sens_spec_results} provides out-of-sample sensitivity and specificity values for the adaptive tests from a subset of maxIPP values shown in Figure \ref{fig:Utility_diffs_train_test}, along with adaptive tests calibrated using utility functions with $w=0.4$ and $w=0.6$. Increasing $w$ results in higher sensitivity and lower specificity, as expected. For full results on maxIPP 2 to 15, along with these quantities for other types of adaptive tests, see the tables in Section 4 of the Supplementary Material (\cite{Krantsevich2022}).

\begin{table}
\scriptsize
\centering
\caption{Sensitivity and specificity on the test data for five adaptive tests, optimized for $w = 0.4$, $0.5$, $0.6$. The synthetic data for calibrating the adaptive tests was obtained from models fit to the training data only.}
\label{table:oos_sens_spec_results}
\begin{tabular}{ c | cc | cc | cc  } 
 \hline
 \textbf{Number of Items}  &  \textbf{Sensitivity} &  \textbf{Specificity} & \textbf{Sensitivity} &  \textbf{Specificity} & \textbf{Sensitivity} &  \textbf{Specificity} \\
 \hline
    &  \multicolumn{2}{c|}{$w = 0.4$ } &  \multicolumn{2}{c|}{$w = 0.5$ }  & \multicolumn{2}{c}{$w = 0.6$ }  \\
\hline
3 & 0.396 & 0.841 & 0.778 & 0.522 & 0.882 & 0.355 \\ 
  6 & 0.507 & 0.816 & 0.771 & 0.587 & 0.931 & 0.300 \\ 
  9 & 0.514 & 0.812 & 0.750 & 0.609 & 0.924 & 0.354 \\ 
  12 & 0.528 & 0.803 & 0.757 & 0.600 & 0.931 & 0.362 \\ 
  15 & 0.528 & 0.803 & 0.757 & 0.600 & 0.931 & 0.372 \\ 
   \hline
\end{tabular}
\end{table}

Finally, we use the holdout set to show how specifying a particular target population can improve sensitivity, specificity, or overall utility when building adaptive tests. The two target populations under consideration are ``All Youth'' and ``Ages 15+''. Table \ref{table:age_counts} shows the number of participants in each of the age groups from our data in both the training and testing sets.
\begin{table}
\caption{Counts of participants in each age group in the training and testing sets.}
\label{table:age_counts}
\centering
\begin{tabular}{  c c  c c }
\hline
 \textbf{Data} & \textbf{Ages 8-14} &  \textbf{Ages 15+}  &  \textbf{Total}    \\ 
\hline
Training Set  & 2297 (82.4\%) &  490 (17.6\%) & 2787 (100\%) \\
Testing Set  &  898 (75.8\%) & 287 (24.2\%) &1185 (100\%) \\
\hline
\end{tabular}
\end{table}
For the adaptive test with target population ``All Youth'', we fit the Gaussian copula factor model and logistic XBART model to the entire training data and obtained synthetic data using these models, which was then used for calibrating the tree-based adaptive test. For the adaptive test with target population ``Ages 15+'',  we used the same models fit to the entire population, but drew synthetic data from the group of youth ages 15 and older using the conditional predictive distribution $f(\tilde{\mathrm{x}}, \tilde{y}\mid \mathrm{x}_{1:n}, y_{1:n}, \text{Age} \geq 15)$. We then calibrated a tree-based adaptive test to this synthetic data. This process was repeated for maxIPP $=2$ to $15$, leaving a total of 28 regression trees. We computed the optimal cutoffs that maximized the utility function (\ref{eqn:expected_utility_intuitive}) for $w=0.6$. After calibrating the trees and computing the optimal cutoffs (using the training data only) to obtain 28 tests, both sets of adaptive tests were then deployed to predict ``at-risk'' status on youth ages 15 and older in the testing set, and sensitivity, specificity, and utility for this group were computed for each of the 28 tests. We chose a value of $0.6$ for this analysis, because we are targeting a group of older youth that have been shown to receive positive treatment effects from the secondary prevention counseling program (see \cite{Katz2021}). For this group it is more important to not miss the youth that are at the highest risk, than to prevent ``not-at-risk'' youth from mistakenly receiving the intervention. 

\begin{figure}
	\includegraphics[width=0.6\linewidth]{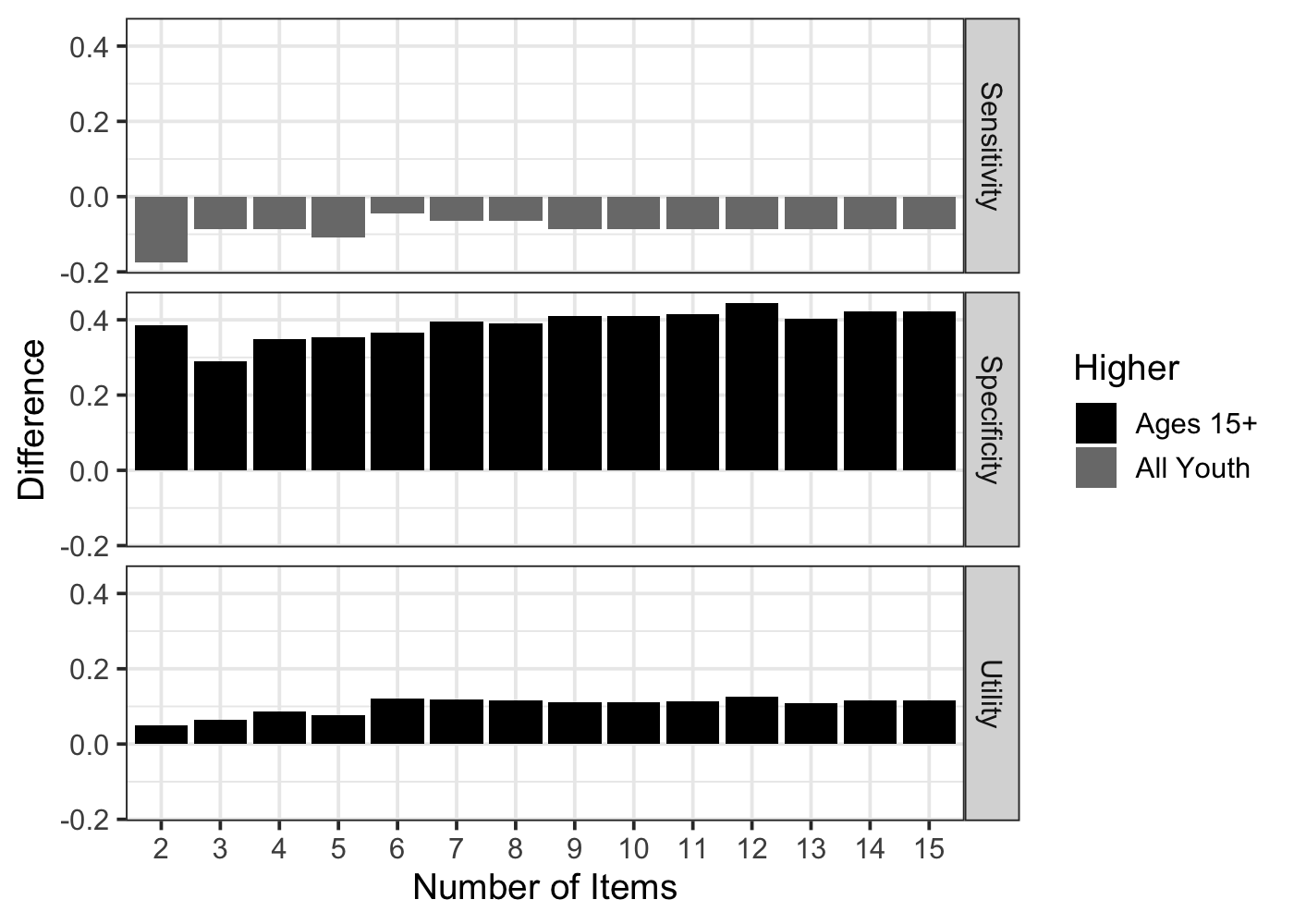}
	\caption{Differences in sensitivity, specificity, and utility for youth ages 15+  in the testing data, between two adaptive tests ($\gamma_{\text{All Youth}}$ and $\gamma_{\text{Ages 15+}}$) created using training data. The adaptive test $\gamma_{\text{All Youth}}$ is designed to approximately optimize expected utility for all youth, and $\gamma_{\text{Ages 15+}}$ for youth ages 15+. The bar height in the upper plot is $\text{Sensitivity}(\gamma_{\text{Ages 15+}}) - \text{Sensitivity}(\gamma_{\text{All Youth}})$ computed on youth ages 15+ in the testing data, and similarly for specificity and utility.}
	\label{fig:subgroup_differences}
\end{figure}

Figure \ref{fig:subgroup_differences} shows the differences in empirical out-of-sample sensitivity, specificity, and overall utility between the adaptive tests calibrated to the two different populations, for each value of maxIPP; the absolute quantities are given in Table \ref{table:subgroup_results}. The adaptive tests optimized for ``All Youth'' with $w=0.6$ are not appropriate for this particular subpopulation, because those questions indicate that all of the youth ages 15+ in the test set are ``at-risk'' (leading to 0 specificity, which clearly is unacceptably low). Trying to increase sensitivity for the entire population results in items that are uninformative for the older youth. When we calibrate the adaptive test specifically to this subgroup, we sacrifice only a small amount of sensitivity for huge gains in specificity.



\begin{table}
\scriptsize
\centering
\caption{Specificity, sensitivity, and utility with $w=0.65$ on youth ages 15 and older from the testing data, for adaptive tests calibrated on two different target populations in the training data: youth ages 15 and older, and all youth. ``Target Population'' shows the population from which synthetic data were obtained for calibrating the test.}
\label{table:subgroup_results}
\begin{tabular}{ c | c c | c c | c c } 
 \hline
 \textbf{Target Population}  & \textbf{Age} $\mathbf{\geq 15}$ & \textbf{All Youth} & \textbf{Age} $\mathbf{\geq 15}$ & \textbf{All Youth} &  \textbf{Age} $\mathbf{\geq 15}$ & \textbf{All Youth} \\
 \hline
   maxIPP & \multicolumn{2}{c |}{Sensitivity} &  \multicolumn{2}{c|}{Specificity } &  \multicolumn{2}{c}{Utility} \\
  \hline
2 & 0.826 & 1.000 & 0.386 & 0.000 & 0.650 & 0.600 \\ 
  3 & 0.913 & 1.000 & 0.290 & 0.000 & 0.664 & 0.600 \\ 
  4 & 0.913 & 1.000 & 0.349 & 0.000 & 0.687 & 0.600 \\ 
  5 & 0.891 & 1.000 & 0.353 & 0.000 & 0.676 & 0.600 \\ 
  6 & 0.957 & 1.000 & 0.365 & 0.000 & 0.720 & 0.600 \\ 
  7 & 0.935 & 1.000 & 0.394 & 0.000 & 0.719 & 0.600 \\ 
  8 & 0.935 & 1.000 & 0.390 & 0.000 & 0.717 & 0.600 \\ 
  9 & 0.913 & 1.000 & 0.411 & 0.000 & 0.712 & 0.600 \\ 
  10 & 0.913 & 1.000 & 0.411 & 0.000 & 0.712 & 0.600 \\ 
  11 & 0.913 & 1.000 & 0.415 & 0.000 & 0.714 & 0.600 \\ 
  12 & 0.913 & 1.000 & 0.444 & 0.000 & 0.725 & 0.600 \\ 
  13 & 0.913 & 1.000 & 0.402 & 0.000 & 0.709 & 0.600 \\ 
  14 & 0.913 & 1.000 & 0.423 & 0.000 & 0.717 & 0.600 \\ 
  15 & 0.913 & 1.000 & 0.423 & 0.000 & 0.717 & 0.600 \\ 
   \hline
\end{tabular}
\end{table}

The improvement by focusing the test to a specific group is an important finding related to \emph{focused deterrence} and \emph{multiple gating}. Focused deterrence implies introducing interventions specific to the group where they will be deployed; multiple gating means targeting youth for secondary prevention programs who are at the highest risk of the delinquent behavior and living within the highest risk neighborhoods (\cite{Katz2021}). Both of these methodologies are important aspects of successful community-based crime prevention programs (\cite{Abt2018}, \cite{Katz2021}), and using accurate screening instruments for the population where an intervention will be introduced is critical to their successful implementation.

While the lack of specificity on the older group of youth using an adaptive test calibrated to all youth is alarming, this highlights the importance of using adaptive tests designed specifically for the group on which they will be deployed. All adaptive tests that are created using a machine learning (ML) algorithm, such as CART, do so by heuristically optimizing a given criterion over a specific dataset. This may have unintended consequences when the data for which the test was optimized differs in distribution to the specific group on which the screening test will be deployed. 

The benefit of our proposed method for obtaining the adaptive test (chosen to optimize the criteria in our Bayesian decision theory evaluation framework), is that these choices are directly placed in front of the screening test designer when the adaptive test is created. One must think critically about the target population for which the test is optimized, and the utility function being optimized--these are decisions that are inherently made in other tree-based adaptive test procedures, but under the hood. 

A further benefit is that data from a larger population can be used to adapt a screening test to a subpopulation where fewer data are available. We borrow information from the whole population when fitting the GCFM model, but sample from the subpopulation of older youth using the conditional posterior predictive distribution from that fitted model. This conditional sample is then used for calibrating the adaptive test to the subgroup. This is an unusual and exciting example of transfer learning---utilizing the information that an ML algorithm obtains from larger datasets when applying the algorithm in service of a slightly different problem where fewer data are available.

\section{Discussion}
\label{section:Discussion}
From a practical perspective, the upshot of our analysis is highly encouraging: a much shorter assessment can be given that will nearly match the predictive accuracy (as characterized by the weighted sensitivity and specificity) of the much longer original assessment. Specifically, we were able to design adaptive tests of varying lengths for the target population of youth ages 15 and older, living in 5 of the poorest and most violent cities in Honduras. Out-of-sample sensitivity over 0.9 and specificity over 0.4 was achieved for an adaptive test that uses only 9 items. This is an increase in specificity of 0.4 over an adaptive test optimized to youth of all ages together. If a more convenient screening tool leads to more individuals being screened, limited crime mitigation resources can be more smartly employed. 

However, precisely because the stakes are so high, circumspection is in order, guided by a ``first do no harm'' ethos. Accordingly, we conclude with an examination of potential pitfalls of our proposed method. The importance of such considerations have recently been emphasized under the broad heading of ``ethical AI'' (artificial intelligence) (cf. \cite{Johndrow2019} and \cite{Chouldechova2020}). 
Two main concerns include disparate impacts on particular subpopulations, and the difficulty in interpreting or interrogating automated decisions from sophisticated data-driven algorithms.

\subsection{Disparate impact}
\label{section:disparate_impact}
Biased training data can result in risk assessment tools that produce unethical or unfair decisions for particular groups of people, in domains such as criminal justice (\cite{Chouldechova2020}, \cite{Chouldechova2017}, \cite{Eckhouse2018}) and child welfare (\cite{Chouldechova2018}). For example, historical data may unfairly indicate that a certain racial group is at higher risk of re-arrest, simply due to more aggressive policing in their neighborhoods; a statistical model trained on this type of historically biased data will produce unethical decisions on important questions like pre-trial release.  Similarly, our method is only as unbiased as the data used to train the model. In our particular case, the outcome used in the IMC data is self-reported; unlike in United States recidivism data, for which ``re-arrest'' is an inaccurate and racially-biased proxy for ``re-offense'' (see \cite{Johndrow2019}), the delinquency data on the IMC is based on the individual youth self-reporting whether they engage in the behavior, as opposed to school or law enforcement records that may be biased by historical law enforcement patterns. While our assessment would disadvantage a group of youth who were systematically dishonest in their self-reported violent behavior, and it is possible that there may be such a group, such patterns in the youth represented in the IMC have thus far not been observed; the scales used in the IMC were chosen for their efficacy, internal validity and reliability (\cite{Katz2021}). 

The nature of historically advantaged or disadvantaged groups also differs: the youth for whom the current application is intended are fairly homogeneous. These youth are of the same race and ethnicity and experience similar levels of poverty, living in the poorest neighborhoods within the five most dangerous and violent cities in Honduras, which is itself one of the most violent countries in the world. While ethnic minority groups live in parts of rural Honduras, this paper has been written for the scope of application in five particular urban neighborhoods under consideration. 


Although our algorithm is unlikely to result in disparate impacts among racial groups in these neighborhoods (simply due to lack of heterogeneity), there is a possibility for differential impact by age, and possibly other features like gender or religion. In \cite{Katz2021}, positive treatment effects from the secondary prevention program were observed for older youth (divided at age 14 and older), whereas mixed treatment effects were observed for the younger group. This highlights the importance of careful selection of the weight $w$ in designing the adaptive test. As a concrete example, in the randomized controlled trial (RCT) which continued after the initial IMC data collection (\cite{Katz2021}), services were given to 994 youth deemed to be ``at-risk'', out of 4495 screened. Supposing that 994 of the 4495 screened youth were truly ``at-risk'', a decrease in sensitivity of 5\% would result in ~50 more ``at-risk'' youth being denied the intervention, whereas a 5\% rise in specificity would result in 175 more ``not-at-risk'' youth being prevented from incorrectly receiving the intervention. An adaptive test that trades this increased specificity for decreased sensitivity may be acceptable within a younger group, but not for an older one. Similarly, harmful consequences can arise from a shift in the target population between test creation and deployment. An adaptive test that optimizes utility for youth over a large age range (e.g., 8-17) may not have acceptable accuracy for youth within a more specific age group; indeed, this was the case for youth ages 15+ (see Section \ref{section:Out of sample validation}).

To summarize, the possibility for disparate impact using our proposed method, as with most automated decision making via ML algorithms, hinges on whether or not particular subpopulations are given due consideration in the test design process. Attention and care must be given to the selection of the target population and the weight $w$ when optimizing the adaptive test, to ensure the best outcomes for the youth being screened for risk of delinquency.

\subsection{Inscrutability of automated decisions}
Independent from concerns surrounding flawed training data and the differential impacts it creates, the sheer complexity of a data driven risk assessment invites skepticism. Flaws can be hard to identify when the inputs and outputs are high dimensional numerical vectors (\cite{Chouldechova2020}). On this count, we consider our method to be a substantive advance over existing approaches. One, our final risk prediction assessment tool is a single decision tree, which can easily be understood and adapted as needed to reduce potential bias or problematic prediction patterns. For example, if a particular item results in lower predicted risk probability based on behavior that is believed to increase it, that item can be excluded from the item pool and the decision tree re-calibrated to the remaining items. Two, the inputs to our method are transparent -- a utility function, a target population, and a set of candidate instruments generated by a heuristic. Sensitivity to these choices can and ought to be investigated; the execution of such comparisons is precisely what our novel decision theory framework facilitates. Although the process is quite involved, its transparency and flexibility should make it {\em less} prone to unanticipated flaws than ad-hoc methods of abridging screening tests, whether data-driven or human guided.


%
%
To emphasize, while particular choices for each of these steps were presented in our analysis of the Honduras data, many other choices are possible. For example, the specific value of $w$ in the utility function can be chosen based on whether specificity or sensitivity is more important; or, another utility function involving other classification metrics can be chosen. The target population can be specified as youth of a particular age, neighborhood, gender, school, or any other subpopulation for which a specific screening instrument may be useful, as long as some data for this target population are available. And while we have focused on tree-based adaptive tests relying on the CART algorithm in this paper, one can utilize other tree-growing algorithms for populating the action space, or compare IRT-based adaptive tests as well. The framework itself is generic, in the sense that once a practitioner has chosen a utility function, a target population, and an algorithm for populating the action space, the same procedure can be applied to understand the trade-offs of shortening the exam to different lengths, or of making a different choice at one of the three steps. 

These choices should be made carefully by policy-makers and local stakeholders, aided by researchers who can explain the trade-offs associated with one decision versus another. Researchers can provide insight via the utility plots, or similar plots created for uncertainty quantification of sensitivity or specificity at the relative or absolute level. Local-stakeholders and policy-makers can assess which outcomes are most important for the group being screened in their specific application. These groups working in concert should adjust the assessment to accommodate desired levels of sensitivity and specificity for the particular population in which it will be deployed, as much as possible considering practical limitations (e.g. counselor availability in our application).

\newpage


\begin{appendix}

\section{Integrating over the target population}
\label{Appendix:Integrating}
Our process for obtaining synthetic data $\{\tilde{\mathrm{x}}_{ij}, \tilde{p}_{ij}, \tilde{y}_{ij}\mid \theta^{(j)}\}$ from the conditional predictive distribution can be summarized as follows:
\begin{quote}
\normalsize
\begin{enumerate}
\item Fit a Gaussian copula factor model with parameters $\theta_\mathrm{X}$ to item response data $\mathrm{x}_{1:n}$. 
\item Fit a multinomial logistic XBART model with parameters $\theta_Y$  to item response/risk status data $(\mathrm{x}, y)_{1:n}$.
\item Fixing the $j^{th}$ posterior draw of model parameters $\theta_\mathrm{X}^{(j)}$, draw $N$ samples $\{\tilde{\mathrm{x}}_{ij}\}_{i=1}^N$ from the conditional predictive distribution $f(\tilde{\mathrm{x}}\mid \theta_\mathrm{X}^{(j)})$ using the fitted Gaussian copula factor model. 
\item Compute the probability $\tilde{p}_{ij}=\text{Pr}(\tilde{Y}=1\mid \tilde{\mathrm{x}}_{ij}, \theta_Y^{(j)})$ using the $j^{th}$ posterior tree ensemble from the fitted multinomial logistic XBART model.
\item Sample the class label $\tilde{y}_{ij}\sim \text{Bernoulli}(\tilde{p}_{ij})$.
\item Our dataset conditioned on the $j^{th}$ posterior draw $\theta^{(j)}=\{\theta_\mathrm{X}^{(j)},\theta_Y^{(j)}\}$ is $\{\tilde{\mathrm{x}}_{ij}, \tilde{p}_{ij}, \tilde{y}_{ij}\mid \theta^{(j)}\}_{i=1}^N$.
\end{enumerate}
\end{quote}

Additionally, during step (4), we compute the posterior predictive mean probability 
\begin{align*}
\bar{\mathbb{E}}(\tilde{Y}\mid \tilde{\mathrm{x}}_{ij}) = \frac{1}{D}\sum_{j=1}^D \tilde{p}_{ij} \approx \int_{\Theta_Y} \mathbb{E}(\tilde{Y} \mid \tilde{\mathrm{x}}_k, \theta_Y) \pi(\theta_Y \mid \mathrm{x}_{1:n},y_{1:n}) \; d\theta_Y.
\end{align*}
By repeating this process $D$ times, $1\leq j \leq D$, we obtain $D$ population-level samples from our target population. In total, the synthetic data is 
\begin{align*}
\{\tilde{\mathrm{x}}_{k}, \hspace{0.1cm} \tilde{p}_{k}, \hspace{0.1cm}\bar{\mathbb{E}}(\tilde{Y}\mid  \tilde{\mathrm{x}}_{k}), \hspace{0.1cm} \tilde{y}_{k}\}_{k=1}^M, \hspace{1cm} M = N \cdot D.
\end{align*}

We use synthetic data $\{\tilde{\mathrm{x}}_{k}, \bar{\mathbb{E}}(\tilde{Y}\mid  \tilde{\mathrm{x}}_{k})\}_{k=1}^M$ for calibrating the regression tree with $m$ items, $T_m^*$. We use $\{\tilde{\mathrm{x}}_{k}, \tilde{\mathrm{y}}_{k}\}_{k=1}^M$ for choosing the optimal cutoff $C_{T_m^*}$. 

We also use $\{\tilde{\mathrm{x}}_{k}, \tilde{\mathrm{y}}_{k}\}_{k=1}^M$ for doing uncertainty quantification plotting, but broken up into $D$ sample populations as $\{\tilde{\mathrm{x}}_{ij}, \tilde{y}_{ij}\mid \theta^{(j)}\}_{i=1}^N$, $1\leq j\leq D$. For each value of $j$, we compute $\mathbb{E}U_{\theta^{(j)}}(\gamma)$  for both 
\begin{align*}
\gamma_m^*(\cdot) = \text{Thr}_{C_{T_m^*}}(T_m^*(\cdot)) \hspace{0.5cm} \text{and} \hspace{0.5cm} \gamma^* (\cdot)= \text{Thr}_{C^*}(\bar{\mathbb{E}}(\tilde{Y} \mid \cdot))
\end{align*}
using Equation (\ref{eqn:computing_utility_from_sample}). The draws of the differences $\Delta_{\theta^{(j)},m}$ between these utilities are then used for uncertainty quantification of $\Delta_{\theta,m}=\mathbb{E}U_{\theta}(\gamma^*_m) - \mathbb{E}U_{\theta}(\gamma^*)$.

For this paper, we drew $N=1000$ Monte Carlo samples of the form 
\begin{align*}
\{\tilde{\mathrm{x}}_{ij}, \tilde{p}_{ij}, \bar{\mathbb{E}}(\tilde{Y}\mid\tilde{\mathrm{x}}_{ij}), \tilde{y}_{ij}\mid \theta^{(j)}\}_{i=1}^N
\end{align*}
for each of $D=1000$ posterior parameter draws $\theta^{(j)}$, $1\leq j \leq 1000$. We drew another 100,000 synthetic data from the same fitted models for the pruning step from Section \ref{section:Populating_the_action_space}.

\section{Details of the maxIPP algorithm}
\label{Appendix:maxIPP criterion}

We propose a variation on the popular CART algorithm for obtaining an approximately optimal tree-based adaptive test that contains at most $m$ items.
 Presuming that item responses can be stored for future splits, the maxIPP of a tree-based adaptive test is precisely the maximum number of questions any participant will answer. The maxIPP characteristic is similar to maximum depth; to see the distinction, consider the tree in the right of Figure \ref{fig:CART}, which has a maximum depth of 3, but a maxIPP of 2.

For each value of maxIPP $=m$, we use an adapted version of the CART algorithm to obtain an approximately optimal tree. CART consists of a tree growing phase, followed by a tree pruning phase. Our modification uses the usual greedy algorithm (minimizing sum-of-squares) for the growing phase, with a twist: once $m$ unique variables have been used as splitting variables in any particular path, only these same variables are considered as candidates for future splits down this path. This algorithm is implemented as a modification to the \texttt{rpart} package, with \texttt{maxvpp} (the application-agnostic term meaning ``\textbf{max}imum \textbf{v}ariables \textbf{p}er \textbf{p}ath'') available as an option for \texttt{rpart.control}. For the pruning stage, we start at the root tree $T_0$ in the list of subtrees returned by \texttt{rpart}, and for each next tree in the list, compute the reduction in root mean square error (on a holdout set) relative to the previous tree. If this reduction is not above a given threshold\footnote{We found that using $10^{-4}$ for maxIPP $<5$ and $10^{-5}$ for maxIPP between 5 and 15 works well in practice; we did not consider maxIPP values above 15 as they produced very similar results to those near 15.}  for at least 10 consecutive subtrees in the list, we return to the last subtree that met this threshold and call this tree $T_m^*$.

\end{appendix}
 \section*{Acknowledgments}
The authors would like to thank Andrew Herren for helpful feedback during the writing of this paper. The first author would like to thank Nikolay Krantsevich for his unending support during the completion of this project.

 The first author was supported by NSF-DMS Award No. 150264.
 
\begin{supplement}
\stitle{Supplement to ``Bayesian decision theory for tree-based adaptive screening tests with an application to youth delinquency''.}

\sdescription{Derivations of the utility function, further information on IMC data and model specifications, and a comparison of different methods for populating the action space.}
\end{supplement}


\bibliographystyle{imsart-nameyear} 
\bibliography{bibliography}       

\end{document}


\begin{frontmatter}
\title{Supplement to ``Bayesian decision theory for tree-based adaptive screening tests with an application to youth delinquency''}
\runtitle{A sample running head title}

\begin{aug}
\author[A]{\fnms{Chelsea} \snm{Krantsevich}\ead[label=e1]{chelsea.krantsevich@asu.edu}},
\author[A]{\fnms{P. Richard} \snm{Hahn}\ead[label=e2,mark]{prhahn@asu.edu}}
\author[A]{\fnms{Yi} \snm{Zheng}\ead[label=e3,mark]{Yi.Isabel.Zheng@asu.edu}}
\and
\author[B]{\fnms{Charles} \snm{Katz}\ead[label=e4,mark]{CHARLES.KATZ@asu.edu}}
\address[A]{School of Mathematical and Statistical Sciences, Arizona State University, \printead{e1,e2,e3}}

\address[B]{School of Criminology and Criminal Justice, Arizona State University, \printead{e4}}
\end{aug}

\end{frontmatter}

\section{Deriving the utility function}
\label{Appendix:Deriving_utility}
Formally, we define our utility function $U$ as
\begin{eqnarray*}
U(\gamma(\mathrm{x}),y) &= \begin{cases} U_0 &\mbox{if } y = 0, \gamma(\mathrm{x}) = 0 \\
U_1 & \mbox{if } y = 1, \gamma(\mathrm{x}) = 1 \\
0 & \mbox{otherwise}\end{cases},
\end{eqnarray*}
where 
\begin{align*}
U_0=\frac{1-w}{\text{Pr}(Y=0)}, \hspace{1cm} U_1=\frac{w}{\text{Pr}(Y=1)}.
\end{align*}

Our expected utility over the target population is
\begin{eqnarray*}
\mathbb{E}U(\gamma)&=&\mathbb{E}[U(\gamma(\mathrm{X}),Y)] \\
&=& \mathbb{E}\left[ \frac{w}{\text{Pr}(Y=1)} \cdot \mathbbm{1}(\gamma(\mathrm{X})=1, Y=1) + \frac{1-w}{\text{Pr}(Y=0)} \cdot \mathbbm{1}(\gamma(\mathrm{X})=0, Y=0)\right]\\
&=&w\cdot\frac{\text{Pr}(\gamma(\mathrm{X})=1, Y=1)}{\text{Pr}(Y=1)}  + (1-w)\cdot \frac{\text{Pr}(\gamma(\mathrm{X})=0, Y=0)}{\text{Pr}(Y=0)} \\
&=&w\cdot \text{Sensitivity}(\gamma) + (1-w)\cdot \text{Specificity}(\gamma).
\end{eqnarray*}

So, the optimal action is
\begin{equation*}
\gamma* = \text{argmax } \mathbb{E}U(\gamma) = \text{argmax } \left\{w\cdot\text{Sensitivity}(\gamma) + (1-w)\cdot\text{Specificity}(\gamma) \right\}.
\end{equation*}

\section{Table on IMC Data}
\label{Appendix:IMC Tables}

Table \ref{table:IMC_factors} provides a list of risk and protective factors measured by the Instrumento de Medicion de Comportamientos (IMC). Data using this instrument were used for obtaining the results in Section 5 of the paper.

\begin{table}
\caption{Risk and protective factors measured by the IMC.}
\label{table:IMC_factors}
\begin{tabular}{lll}
\hline
\multicolumn{1}{l}{\textbf{Domain}}& \multicolumn{1}{l}{\textbf{Risk Factors}} &\multicolumn{1}{l}{\textbf{Protective Factors}} \\
\hline
Community   & Transitions and mobility & Rewards for prosocial involvement\\
& Low neighborhood attachment  & Opportunities for prosocial involvement \\
& Community disorganization & \\
& Laws and norms favorable to drug use & \\
& Perceived availability of drugs  \\
\hline
Family  & Family history of antisocial behavior & Attachment \\
& Parental attitudes favorable towards drug use & Opportunities for prosocial involvement\\
&Poor family management & Rewards for prosocial involvement\\
&Family conflict & \\
& Weak parental supervision & \\
& Family gang influence & \\
\hline
School  & Academic failure & Opportunities for prosocial involvement\\
& Low commitment to school & Rewards for prosocial involvement\\
\hline
Peer/Individual  & Rebelliousness & Belief in the moral order\\
& Rewards for antisocial involvement & Rewards for prosocial involvement\\
& Favorable attitudes towards drug use & Interaction with prosocial peers\\
& Favorable attitudes towards antisocial behavior \hspace{.2in} & Social skills\\
& Perceived risks of drug use & \\
& Friends' use of drugs & \\
& Interaction with antisocial peers & \\
& Intentions to use & \\
& Antisocial tendencies & \\
& Critical life events & \\
& Impulsive risk taking & \\
& Neutralization of guilt & \\
& Negative peer influence & \\
& Peer delinquency & \\
\hline
\end{tabular}
\normalsize
\end{table}

\section{Model Specifications}
\subsection{Item responses}
\label{Appendix:BFA model}
As described in \cite{Carvalho2006}, in a $k$-dimensional Gaussian factor model, the $i^{th}$ observation of a $p\times 1$ random vector $\mathrm{z}$ can be represented as
\begin{equation*}
\mathrm{z}_i = \mathbf{\Lambda}\mathrm{f}_i + \mathrm{\nu}_i,
\end{equation*}
where $\mathbf{\Lambda}$ is a $p\times k$ matrix of factor loadings, $\mathrm{f}_i$ is a $k\times 1$ vector of factor scores with $\mathrm{f}_i \sim N(0, \mathbf{I})$, and $\boldsymbol{\nu}_i$ is a $p\times 1$ noise vector, with $\boldsymbol{\nu}_i \sim N(0, \boldsymbol{\Psi})$, $\boldsymbol{\Psi} = \text{diag}(\Psi_1, \dots, \Psi_p)$. Under these assumptions, $\mathrm{z}\sim N(0, \mathbf{\Lambda} \mathbf{\Lambda}^{T}+\boldsymbol{\Psi})$. 

Intuitively, a copula is a joint distribution function that allows for separation of the marginals from the dependence structure; the copula completely describes all dependencies among variables. A joint distribution $F$ has a Gaussian copula if it can be written as 
\begin{equation*}
F(X_1, X_2,\dots, X_p) = \Phi_p(\Phi^{-1}(F_1(X_1)),\Phi^{-1}(F_2(X_2)),\dots, \Phi^{-1}(F_p(X_p))\mid \mathcal{C}),
\end{equation*}
where $\Phi_p$ is the $p$-dimensional multivariate Gaussian CDF with correlation matrix $\mathcal{C}$, and $\Phi^{-1}$ is the inverse Gaussian CDF.

The Gaussian copula factor model starts by assigning the latent variable $\mathrm{z}$ a $k$-dimensional Gaussian factor model: $\mathrm{f}_i \sim N(0, \mathbf{I})$, $\mathrm{z}_i\mid \mathrm{f}_i \sim N(\mathbf{\Lambda}\mathrm{f}_i, \mathbf{I})$. We then define $\mathrm{x}$ as 
\begin{equation*}
x_{ir}=F_r^{-1}\left(\Phi\left(\frac{z_{ir}}{\sqrt{1+\sum_{t=1}^k \lambda_{rt}^2}}\right)\right),
\end{equation*}
where $F_r^{-1}(t)=\text{inf}\{t: F_r(x)\geq t, x\in \mathbb{R}\}$ is the pseudo-inverse of $F_r, 1\leq r \leq p$. By making this specification, $F(\mathrm{x})$ has a Gaussian copula with covariance matrix $\mathbf{\Lambda}\mathbf{\Lambda}^T + \mathbf{I}$, and marginals $F_1, F_2, \dots, F_p$.

The \texttt{bfa} R package presented in \cite{Murray2013} fits a Gaussian copula factor model to data using a parameter-expanded Gibbs sampling scheme. Their method allows for inference on joint distributions of mixed continuous and discrete variables, which is necessary for modeling the joint distribution of the item responses and demographic variables from the IMC data. We fit the Gaussian copula factor model to item response and demographic data from the target population in the IMC data, then obtain samples $\{\mathrm{\tilde{x_i}}\}_{i=1}^k$ using the predictive distribution of the fitted model.

\subsection{Risk prediction}
\label{Appendix:XBART model}
In the log-linear Accelerated Bayesian Additive Regression Trees (XBART) model, the probability of observing class $s$ given covariate $\mathrm{x}_i$ in a setting with $c$ classes follows a logistic specification
\begin{equation*}
\pi_s(\mathrm{x}_i) = \frac{h^{(s)}(\mathrm{x}_i)}{\sum_{t=1}^c h^{(t)}(\mathrm{x}_i)}.
\end{equation*}

Following \cite{Murray2020},  $\log[h^{(s)}(\mathrm{x}_i)] = \sum_{l=1}^L g(\mathrm{x}_i, T_l^{(s)}, \mu_l^{(s)})$ is given a sum of trees representation, where $g(\mathrm{x}_i, T_l^{(s)}, \mu_l^{(s)})$ is a tree with splits given by $T_l^{(s)}$ and leaf mean parameter $\mu_l^{(s)}$. This yields
\begin{equation*}
\pi_s(\mathrm{x}_i) = \frac{\exp [\sum_{l=1}^L g(\mathrm{x}_i, T_l^{(s)}, \mu_l^{(s)})]}{\sum_{t=1}^c \exp[\sum_{l=1}^L g(\mathrm{x}_i, T_l^{(t)}, \mu_l^{(t)})]}.
\end{equation*}
In our application, we utilize this multinomial logistic XBART model with $c=2$ classes to predict the probability of being ``at-risk'', given item responses $\mathrm{x}_i$, as $\bar{\mathbb{E}}(\tilde{Y}\mid \mathrm{x}_i) = \pi_{s=1}(\mathrm{x}_i)$.

\section{Changing the algorithm to populate the action space}
\label{Appendix:Changing action space}

In the paper we present one method for populating the action space with a single adaptive test of length $m$. First, we calibrate a regression tree $T^*_m$ to synthetic data sampled from the posterior predictive distribution, using the maxIPP algorithm with maxIPP $=m$. Then we choose a threshold which is optimized relative to $T^*_m$. This is only one possible heuristic for obtaining a tree-based adaptive test with at most $m$ questions. We can change the tree-growing algorithm, the data the algorithm is applied to, the way the threshold is chosen; or, we can calibrate a classification tree directly instead of using a two-stage regression tree + cutoff approach. Here we present results for several such alternatives.

First, we obtain different adaptive tests varying the parameters above, using the entire IMC data set. We compare two stopping criteria for growing the regression tree: 1) growing to a maximum depth; 2) growing very deep using maxIPP then pruning back (proposed method, described in Appendix B of the paper). We furthermore compare these tree-growing algorithms applied to synthetic data generated via two different processes: 

\begin{quote}
\normalsize
\begin{enumerate}
\item Item responses generated via local perturbations (``Perturb'') and ``at-risk'' probabilities obtained using a Random Forest model (``RF''), as in \cite{Gibbons2013}.
\item Item responses sampled from a Gaussian copula factor model (``GCFM'') and ``at-risk'' probabilities obtained using a logistic Accelerated Bayesian Additive Regression Trees (``XBART'') model. This is our proposed method, described in Sections 3.2.2 and 3.2.3 of the main paper, with data notated $\{\tilde{\mathrm{x}}_k, \bar{\mathbb{E}}(\tilde{Y}\mid\tilde{\mathrm{x}}_k)\}$, $1\leq k \leq M$.
\end{enumerate}
\end{quote}

For each of these two regression trees, we obtain an optimal cutoff using data $\{\tilde{\mathrm{x}}_{k}, \tilde{y}_k\}_{k=1}^M$, via the method described in Section 3.2.3. 

We also consider the simpler approach of calibrating a classification tree (via maximum depth) that predicts ``at-risk'' status directly, rather than the regression tree + cutoff approach. We calibrate three classification trees using the following synthetic datasets:
\begin{quote}
\normalsize
\begin{enumerate}
\item Item responses generated via local perturbations (``Perturb'') and risk classes (``at-risk'' = 1, ``not-at-risk'' = 0) obtained using RF. This data uses the same item responses and the same fitted model as above, but extracting class predictions rather than probability of being in the ``at-risk'' class.
\item Synthetic item responses and classes $\{\tilde{\mathrm{x}}_k, \tilde{y}_k\}$, $1\leq k \leq M$, described in Section 3.2.2 and Appendix A of the main paper. 
\item Synthetic item responses and utility-based synthetic classes $\{\tilde{\mathrm{x}}_k, \gamma_k^*\}$, $1\leq k \leq M$, described below.
\end{enumerate}
\end{quote}

A reviewer suggested classification tree method (3). Unlike the first two synthetic datasets, which do not incorporate the utility function at all, this method still approximately maximizes the utility function (subject to the number of items constraint), and allows for fine-tuning $w$ based on desired levels of sensitivity and specificity. 

The synthetic class outcomes $\gamma_k^*$ are defined as:
\begin{eqnarray*}
\gamma_k^* = \gamma(\tilde{\mathrm{x_k}}) &= \begin{cases} 1 &\mbox{if } \bar{\mathbb{E}}(\tilde{Y}\mid \tilde{\mathrm{x}}_k) \geq \frac{U_0}{U_0+U_1} \\
0 & \mbox{otherwise}\end{cases},
\end{eqnarray*}
where $U_0$ and $U_1$ are defined in Section \ref{Appendix:Deriving_utility} of this Supplement. If one was not required to use a decision tree for classifying participants as ``at-risk'' or ``not-at-risk'', this class assignment would maximize the expected posterior utility point-wise. Thus, a classification tree calibrated to this third synthetic dataset will approximately optimize the expected utility function.

In summary, the settings compared for different adaptive tests are:
\begin{quote}
\normalsize
\begin{enumerate}
\item Regression Tree + Cutoff Methods
\begin{enumerate}
\item Stopping criterion: 
\begin{itemize}
 \item maxIPP \& pruning (maxIPP)
 \item  maximum depth (maxDepth)
\end{itemize} 
\item Data for calibrating (item responses + predicted probabilities): 
\begin{itemize}
\item  local perturbations + Random Forest (Perturb + RF)
\item Gaussian copula factor model + accelerated Bayesian additive regression trees  (GCFM + XBART)
\end{itemize} 
\end{enumerate}
\item Classification Tree Method
\begin{enumerate}
\item Data for calibrating (item responses + predicted classes): 
\begin{itemize}
\item  Perturb + RF
\item  GCFM + XBART
\item GCFM + utility-based outcomes (GCFM + Utility)
\end{itemize}
\end{enumerate}
\end{enumerate}
\end{quote}
In total this leaves four regression tree methods and three classification tree methods. For $\gamma$ being each of the seven methods, we computed the expected utility draw $\mathbb{E}U_{\theta^{(j)}}(\gamma)$ over the $j^{th}$ sample population $\{\tilde{\mathrm{x}}_{ij}, \tilde{y}_{ij}\}_{i=1}^N$ using Eqn (4). We similarly computed the $j^{th}$ draw of expected utility of the non-shortened assessment $\gamma^*(\cdot) = \text{Thr}_{C^*}(\bar{\mathbb{E}}(\tilde{Y}\mid\cdot))$, and the difference between the two expected utilities is $\Delta_{\theta^{(j)},m}$.

\begin{figure}
	\includegraphics[width=\linewidth]{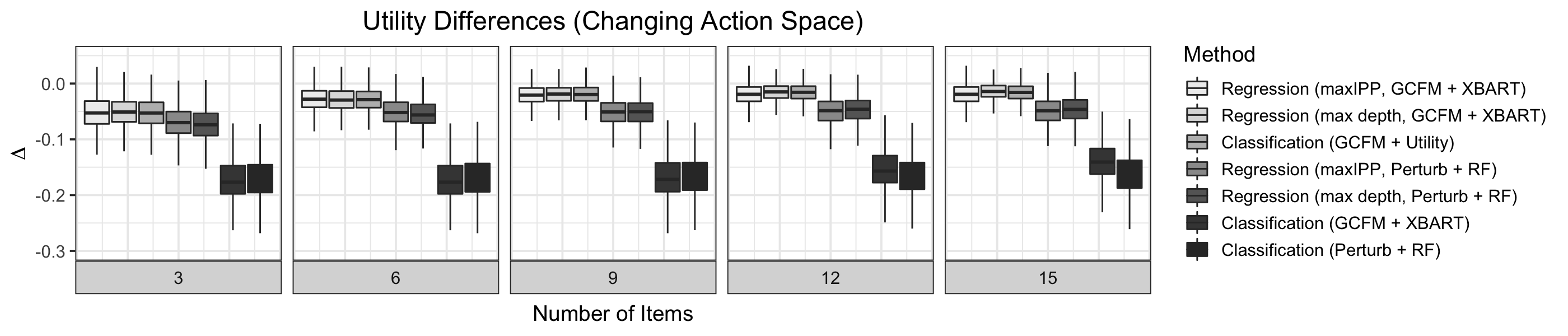}
	\caption{Utility difference plots when we change the way of populating the action space.}
	\label{fig:util_diffs_competitors}
\end{figure}

Figure \ref{fig:util_diffs_competitors} shows the boxplots representing the distribution of $\Delta_{\theta,m}$ for each of these seven methods, for a utility function weight of $w=0.5$. The maxIPP/maximum depth criteria are grouped as ``Number of Items''.  

First, we emphasize that Figure \ref{fig:util_diffs_competitors} is not intended to demonstrate the ultimate superiority of any method, but rather to assess which method best approximates our implementation of an optimal (in terms of expected utility) screening instrument that uses all items. We make a direct comparison of these methods on out-of-sample data shortly.

The most striking result in Figure \ref{fig:util_diffs_competitors} is the superiority of the utility-based methods (all 4 regression-based methods, and the utility-based classification method) at reproducing the utility of the full-item instrument. This is unsurprising because the other two classification methods were not created with the utility function in mind; they were simply calibrated to synthetic data obtained from models fit to the IMC data. The IMC data is highly imbalanced, so these two classification trees tend to predict almost all youth to be ``not-at-risk'', and they have low expected utility compared to the non-sparse action.

The maxIPP and maximum depth stopping criteria produce similar results. We expect there to be a greater difference using maxIPP for an application in which multiple strata of the same question lead to to substantially different outcomes, whereas in this particular application, we feel that differences in item responses are mainly important in the ``high-low'' sense. However, the maxIPP process seems to do at least as well as maximum depth.

We also derived out of sample sensitivity and specificity for all 7 of these methods in order to do a fair comparison between them. The results are shown in the tables, at the end of this Appendix, one table per utility function (parameterized by $w$).

Sensitivity is quite poor for the non-utility based classification methods, presumably because the IMC data is highly imbalanced. As a consequence, synthetic data used from models fitted to this imbalanced data also have very few ``at-risk'' predictions, and so these classification trees in turn make very few ``at-risk'' predictions. We note that the method Classification -- maxDepth -- Perturb + RF was the method used for designing the tree-based adaptive tests in \cite{Gibbons2016}, according to our personal communication with the authors (\cite{Gibbons2019}). While their method seems to work well for problems such as screening for Major Depressive Disorder, in our application, which has extremely imbalanced class outcomes, it performs quite poorly. 

These results highlight the benefit of optimizing the adaptive test to the utility function, either using a regression tree and a separately optimized cutoff, or using a classification tree calibrated to utility-based class labels. In both of these methods, we are able to make finer adjustments to the screening procedure and balance sensitivity and specificity to be in line with desired ranges for a particular group. 

For all of the utility-based methods (all regression methods and the utility-based cutoff method), increasing $w$ improves the sensitivity and decreases the specificity, as expected. The regression methods using GCFM + XBART and the utility based classification methods are two different ways of obtaining an adaptive test that approximates the same utility function using the same fitted models, the only difference being whether the utility is optimized point-wise or globally.

We see three advantages of our proposed two-step (global optimization) method:

\begin{quote}
\normalsize
\begin{enumerate}
\item We believe the two-step formulation is more intuitive, because the connection to sensitivity and specificity in the quantity used to optimize the threshold (Equation (6) in the main paper) is more direct.  

\item The two-step process allows one to visualize the impact of the weight $w$ via ROC curves, as in Figure 5 from the main paper. Since the predicted probabilities do not change, and the utility function is optimized via the threshold function $\text{Thr}_{C_T}$, one can plot a ROC curve of the predicted probabilities. This allows for directly see how changing $w$ results in different thresholds and corresponding Sensitivity and Specificity.

\item In terms of practical implementation, it is much more efficient to fit only one regression tree and then separately optimize the threshold for many values of $w$, compared to fitting a new classification tree for every new value of $w$ one would like to examine. Since our MCMC sample contains 1 million synthetic data, this is actually a substantial computational speedup when many values of $w$ are under consideration. 

\end{enumerate}
\end{quote}

\newpage 
\footnotesize
\begin{longtable}{ccccccc}
  \hline
Number of Items & Tree Type & Criterion & w & Calibration Data & Sensitivity & Specificity \\ 
  \hline
2 & Classification & maxDepth & - & Perturb + RF & 0.007 & 1.000 \\ 
  2 & Classification & maxDepth & - & GCFM + XBART & 0.000 & 1.000 \\ 
     \hline
  3 & Classification & maxDepth & - & Perturb + RF & 0.007 & 1.000 \\ 
  3 & Classification & maxDepth & - & GCFM + XBART & 0.000 & 1.000 \\ 
     \hline
  4 & Classification & maxDepth & - & Perturb + RF & 0.014 & 1.000 \\ 
  4 & Classification & maxDepth & - & GCFM + XBART & 0.000 & 1.000 \\ 
     \hline
  5 & Classification & maxDepth & - & Perturb + RF & 0.014 & 0.999 \\ 
  5 & Classification & maxDepth & - & GCFM + XBART & 0.000 & 1.000 \\ 
     \hline
  6 & Classification & maxDepth & - & Perturb + RF & 0.028 & 0.997 \\ 
  6 & Classification & maxDepth & - & GCFM + XBART & 0.000 & 1.000 \\ 
     \hline
  7 & Classification & maxDepth & - & Perturb + RF & 0.021 & 0.997 \\ 
  7 & Classification & maxDepth & - & GCFM + XBART & 0.000 & 1.000 \\ 
     \hline
  8 & Classification & maxDepth & - & Perturb + RF & 0.028 & 0.997 \\ 
  8 & Classification & maxDepth & - & GCFM + XBART & 0.000 & 0.999 \\ 
     \hline
  9 & Classification & maxDepth & - & Perturb + RF & 0.021 & 0.997 \\ 
  9 & Classification & maxDepth & - & GCFM + XBART & 0.000 & 0.999 \\ 
     \hline
  10 & Classification & maxDepth & - & Perturb + RF & 0.028 & 0.995 \\ 
  10 & Classification & maxDepth & - & GCFM + XBART & 0.042 & 0.994 \\ 
     \hline
  11 & Classification & maxDepth & - & Perturb + RF & 0.021 & 0.994 \\ 
  11 & Classification & maxDepth & - & GCFM + XBART & 0.049 & 0.992 \\ 
     \hline
  12 & Classification & maxDepth & - & Perturb + RF & 0.021 & 0.995 \\ 
  12 & Classification & maxDepth & - & GCFM + XBART & 0.062 & 0.980 \\ 
     \hline
  13 & Classification & maxDepth & - & Perturb + RF & 0.028 & 0.994 \\ 
  13 & Classification & maxDepth & - & GCFM + XBART & 0.097 & 0.976 \\ 
     \hline
  14 & Classification & maxDepth & - & Perturb + RF & 0.035 & 0.994 \\ 
  14 & Classification & maxDepth & - & GCFM + XBART & 0.132 & 0.967 \\ 
     \hline
  15 & Classification & maxDepth & - & Perturb + RF & 0.035 & 0.992 \\ 
  15 & Classification & maxDepth & - & GCFM + XBART & 0.160 & 0.955 \\ 
   \hline
\end{longtable}
 
  \begin{longtable}{ccccccc}
  \hline
Number of Items & Tree Type & Criterion & w & Calibration Data & Sensitivity & Specificity \\ 
  \hline
  2 & Regression + Cutoff & maxIPP & 0.4 & Perturb + RF & 0.319 & 0.908 \\ 
  2 & Regression + Cutoff & maxDepth & 0.4 & Perturb + RF & 0.319 & 0.908 \\ 
  2 & Regression + Cutoff & maxIPP & 0.4 & GCFM + XBART & 0.340 & 0.857 \\ 
  2 & Regression + Cutoff & maxDepth & 0.4 & GCFM + XBART & 0.167 & 0.940 \\ 
  2 & Classification & maxDepth & 0.4 & GCFM + Utility & 0.167 & 0.940 \\ 
     \hline
  3 & Regression + Cutoff & maxIPP & 0.4 & Perturb + RF & 0.306 & 0.907 \\ 
  3 & Regression + Cutoff & maxDepth & 0.4 & Perturb + RF & 0.389 & 0.885 \\ 
  3 & Regression + Cutoff & maxIPP & 0.4 & GCFM + XBART & 0.396 & 0.841 \\ 
  3 & Regression + Cutoff & maxDepth & 0.4 & GCFM + XBART & 0.312 & 0.891 \\ 
  3 & Classification & maxDepth & 0.4 & GCFM + Utility & 0.326 & 0.879 \\ 
     \hline
  4 & Regression + Cutoff & maxIPP & 0.4 & Perturb + RF & 0.444 & 0.848 \\ 
  4 & Regression + Cutoff & maxDepth & 0.4 & Perturb + RF & 0.444 & 0.848 \\ 
  4 & Regression + Cutoff & maxIPP & 0.4 & GCFM + XBART & 0.424 & 0.841 \\ 
  4 & Regression + Cutoff & maxDepth & 0.4 & GCFM + XBART & 0.354 & 0.872 \\ 
  4 & Classification & maxDepth & 0.4 & GCFM + Utility & 0.354 & 0.860 \\ 
     \hline
  5 & Regression + Cutoff & maxIPP & 0.4 & Perturb + RF & 0.438 & 0.866 \\ 
  5 & Regression + Cutoff & maxDepth & 0.4 & Perturb + RF & 0.382 & 0.888 \\ 
  5 & Regression + Cutoff & maxIPP & 0.4 & GCFM + XBART & 0.465 & 0.834 \\ 
  5 & Regression + Cutoff & maxDepth & 0.4 & GCFM + XBART & 0.438 & 0.833 \\ 
  5 & Classification & maxDepth & 0.4 & GCFM + Utility & 0.417 & 0.861 \\ 
     \hline
  6 & Regression + Cutoff & maxIPP & 0.4 & Perturb + RF & 0.514 & 0.859 \\ 
  6 & Regression + Cutoff & maxDepth & 0.4 & Perturb + RF & 0.472 & 0.875 \\ 
  6 & Regression + Cutoff & maxIPP & 0.4 & GCFM + XBART & 0.507 & 0.816 \\ 
  6 & Regression + Cutoff & maxDepth & 0.4 & GCFM + XBART & 0.535 & 0.797 \\ 
  6 & Classification & maxDepth & 0.4 & GCFM + Utility & 0.438 & 0.853 \\ 
     \hline
  7 & Regression + Cutoff & maxIPP & 0.4 & Perturb + RF & 0.562 & 0.852 \\ 
  7 & Regression + Cutoff & maxDepth & 0.4 & Perturb + RF & 0.521 & 0.865 \\ 
  7 & Regression + Cutoff & maxIPP & 0.4 & GCFM + XBART & 0.500 & 0.823 \\ 
  7 & Regression + Cutoff & maxDepth & 0.4 & GCFM + XBART & 0.486 & 0.826 \\ 
  7 & Classification & maxDepth & 0.4 & GCFM + Utility & 0.500 & 0.835 \\ 
     \hline
  8 & Regression + Cutoff & maxIPP & 0.4 & Perturb + RF & 0.500 & 0.858 \\ 
  8 & Regression + Cutoff & maxDepth & 0.4 & Perturb + RF & 0.451 & 0.869 \\ 
  8 & Regression + Cutoff & maxIPP & 0.4 & GCFM + XBART & 0.569 & 0.780 \\ 
  8 & Regression + Cutoff & maxDepth & 0.4 & GCFM + XBART & 0.535 & 0.796 \\ 
  8 & Classification & maxDepth & 0.4 & GCFM + Utility & 0.507 & 0.834 \\ 
     \hline
  9 & Regression + Cutoff & maxIPP & 0.4 & Perturb + RF & 0.493 & 0.865 \\ 
  9 & Regression + Cutoff & maxDepth & 0.4 & Perturb + RF & 0.479 & 0.866 \\ 
  9 & Regression + Cutoff & maxIPP & 0.4 & GCFM + XBART & 0.514 & 0.812 \\ 
  9 & Regression + Cutoff & maxDepth & 0.4 & GCFM + XBART & 0.486 & 0.827 \\ 
  9 & Classification & maxDepth & 0.4 & GCFM + Utility & 0.507 & 0.828 \\ 
     \hline
  10 & Regression + Cutoff & maxIPP & 0.4 & Perturb + RF & 0.507 & 0.854 \\ 
  10 & Regression + Cutoff & maxDepth & 0.4 & Perturb + RF & 0.521 & 0.850 \\ 
  10 & Regression + Cutoff & maxIPP & 0.4 & GCFM + XBART & 0.528 & 0.799 \\ 
  10 & Regression + Cutoff & maxDepth & 0.4 & GCFM + XBART & 0.528 & 0.790 \\ 
  10 & Classification & maxDepth & 0.4 & GCFM + Utility & 0.535 & 0.828 \\ 
     \hline
  11 & Regression + Cutoff & maxIPP & 0.4 & Perturb + RF & 0.479 & 0.864 \\ 
  11 & Regression + Cutoff & maxDepth & 0.4 & Perturb + RF & 0.479 & 0.861 \\ 
  11 & Regression + Cutoff & maxIPP & 0.4 & GCFM + XBART & 0.528 & 0.803 \\ 
  11 & Regression + Cutoff & maxDepth & 0.4 & GCFM + XBART & 0.549 & 0.795 \\ 
  11 & Classification & maxDepth & 0.4 & GCFM + Utility & 0.535 & 0.824 \\ 
     \hline
  12 & Regression + Cutoff & maxIPP & 0.4 & Perturb + RF & 0.514 & 0.846 \\ 
  12 & Regression + Cutoff & maxDepth & 0.4 & Perturb + RF & 0.521 & 0.841 \\ 
  12 & Regression + Cutoff & maxIPP & 0.4 & GCFM + XBART & 0.528 & 0.803 \\ 
  12 & Regression + Cutoff & maxDepth & 0.4 & GCFM + XBART & 0.521 & 0.803 \\ 
  12 & Classification & maxDepth & 0.4 & GCFM + Utility & 0.521 & 0.815 \\ 
     \hline
  13 & Regression + Cutoff & maxIPP & 0.4 & Perturb + RF & 0.514 & 0.846 \\ 
  13 & Regression + Cutoff & maxDepth & 0.4 & Perturb + RF & 0.556 & 0.821 \\ 
  13 & Regression + Cutoff & maxIPP & 0.4 & GCFM + XBART & 0.528 & 0.803 \\ 
  13 & Regression + Cutoff & maxDepth & 0.4 & GCFM + XBART & 0.528 & 0.807 \\ 
  13 & Classification & maxDepth & 0.4 & GCFM + Utility & 0.521 & 0.835 \\ 
     \hline
  14 & Regression + Cutoff & maxIPP & 0.4 & Perturb + RF & 0.514 & 0.846 \\ 
  14 & Regression + Cutoff & maxDepth & 0.4 & Perturb + RF & 0.549 & 0.819 \\ 
  14 & Regression + Cutoff & maxIPP & 0.4 & GCFM + XBART & 0.528 & 0.803 \\ 
  14 & Regression + Cutoff & maxDepth & 0.4 & GCFM + XBART & 0.542 & 0.785 \\ 
  14 & Classification & maxDepth & 0.4 & GCFM + Utility & 0.549 & 0.823 \\ 
     \hline
  15 & Regression + Cutoff & maxIPP & 0.4 & Perturb + RF & 0.514 & 0.846 \\ 
  15 & Regression + Cutoff & maxDepth & 0.4 & Perturb + RF & 0.479 & 0.823 \\ 
  15 & Regression + Cutoff & maxIPP & 0.4 & GCFM + XBART & 0.528 & 0.803 \\ 
  15 & Regression + Cutoff & maxDepth & 0.4 & GCFM + XBART & 0.535 & 0.795 \\ 
  15 & Classification & maxDepth & 0.4 & GCFM + Utility & 0.562 & 0.822 \\ 
   \hline
\end{longtable}
  
  \begin{longtable}{ccccccc}
  \hline
Number of Items & Tree Type & Criterion & w & Calibration Data & Sensitivity & Specificity \\ 
  \hline
  2 & Regression + Cutoff & maxIPP & 0.5 & Perturb + RF & 0.583 & 0.748 \\ 
  2 & Regression + Cutoff & maxDepth & 0.5 & Perturb + RF & 0.868 & 0.465 \\ 
  2 & Regression + Cutoff & maxIPP & 0.5 & GCFM + XBART & 0.750 & 0.534 \\ 
  2 & Regression + Cutoff & maxDepth & 0.5 & GCFM + XBART & 0.840 & 0.425 \\ 
  2 & Classification & maxDepth & 0.5 & GCFM + Utility & 0.549 & 0.722 \\ 
     \hline
  3 & Regression + Cutoff & maxIPP & 0.5 & Perturb + RF & 0.750 & 0.606 \\ 
  3 & Regression + Cutoff & maxDepth & 0.5 & Perturb + RF & 0.688 & 0.658 \\ 
  3 & Regression + Cutoff & maxIPP & 0.5 & GCFM + XBART & 0.778 & 0.522 \\ 
  3 & Regression + Cutoff & maxDepth & 0.5 & GCFM + XBART & 0.660 & 0.659 \\ 
  3 & Classification & maxDepth & 0.5 & GCFM + Utility & 0.583 & 0.691 \\ 
     \hline
  4 & Regression + Cutoff & maxIPP & 0.5 & Perturb + RF & 0.792 & 0.627 \\ 
  4 & Regression + Cutoff & maxDepth & 0.5 & Perturb + RF & 0.750 & 0.671 \\ 
  4 & Regression + Cutoff & maxIPP & 0.5 & GCFM + XBART & 0.708 & 0.652 \\ 
  4 & Regression + Cutoff & maxDepth & 0.5 & GCFM + XBART & 0.757 & 0.584 \\ 
  4 & Classification & maxDepth & 0.5 & GCFM + Utility & 0.660 & 0.677 \\ 
     \hline
  5 & Regression + Cutoff & maxIPP & 0.5 & Perturb + RF & 0.764 & 0.614 \\ 
  5 & Regression + Cutoff & maxDepth & 0.5 & Perturb + RF & 0.806 & 0.566 \\ 
  5 & Regression + Cutoff & maxIPP & 0.5 & GCFM + XBART & 0.743 & 0.618 \\ 
  5 & Regression + Cutoff & maxDepth & 0.5 & GCFM + XBART & 0.750 & 0.598 \\ 
  5 & Classification & maxDepth & 0.5 & GCFM + Utility & 0.743 & 0.593 \\ 
     \hline
  6 & Regression + Cutoff & maxIPP & 0.5 & Perturb + RF & 0.750 & 0.661 \\ 
  6 & Regression + Cutoff & maxDepth & 0.5 & Perturb + RF & 0.799 & 0.601 \\ 
  6 & Regression + Cutoff & maxIPP & 0.5 & GCFM + XBART & 0.771 & 0.587 \\ 
  6 & Regression + Cutoff & maxDepth & 0.5 & GCFM + XBART & 0.764 & 0.573 \\ 
  6 & Classification & maxDepth & 0.5 & GCFM + Utility & 0.694 & 0.633 \\ 
     \hline
  7 & Regression + Cutoff & maxIPP & 0.5 & Perturb + RF & 0.840 & 0.587 \\ 
  7 & Regression + Cutoff & maxDepth & 0.5 & Perturb + RF & 0.785 & 0.668 \\ 
  7 & Regression + Cutoff & maxIPP & 0.5 & GCFM + XBART & 0.729 & 0.612 \\ 
  7 & Regression + Cutoff & maxDepth & 0.5 & GCFM + XBART & 0.715 & 0.640 \\ 
  7 & Classification & maxDepth & 0.5 & GCFM + Utility & 0.708 & 0.641 \\ 
     \hline
  8 & Regression + Cutoff & maxIPP & 0.5 & Perturb + RF & 0.799 & 0.594 \\ 
  8 & Regression + Cutoff & maxDepth & 0.5 & Perturb + RF & 0.778 & 0.638 \\ 
  8 & Regression + Cutoff & maxIPP & 0.5 & GCFM + XBART & 0.743 & 0.622 \\ 
  8 & Regression + Cutoff & maxDepth & 0.5 & GCFM + XBART & 0.778 & 0.566 \\ 
  8 & Classification & maxDepth & 0.5 & GCFM + Utility & 0.729 & 0.624 \\ 
     \hline
  9 & Regression + Cutoff & maxIPP & 0.5 & Perturb + RF & 0.799 & 0.581 \\ 
  9 & Regression + Cutoff & maxDepth & 0.5 & Perturb + RF & 0.792 & 0.592 \\ 
  9 & Regression + Cutoff & maxIPP & 0.5 & GCFM + XBART & 0.750 & 0.609 \\ 
  9 & Regression + Cutoff & maxDepth & 0.5 & GCFM + XBART & 0.750 & 0.606 \\ 
  9 & Classification & maxDepth & 0.5 & GCFM + Utility & 0.736 & 0.634 \\ 
     \hline
  10 & Regression + Cutoff & maxIPP & 0.5 & Perturb + RF & 0.806 & 0.571 \\ 
  10 & Regression + Cutoff & maxDepth & 0.5 & Perturb + RF & 0.806 & 0.577 \\ 
  10 & Regression + Cutoff & maxIPP & 0.5 & GCFM + XBART & 0.757 & 0.603 \\ 
  10 & Regression + Cutoff & maxDepth & 0.5 & GCFM + XBART & 0.750 & 0.601 \\ 
  10 & Classification & maxDepth & 0.5 & GCFM + Utility & 0.771 & 0.621 \\ 
     \hline
  11 & Regression + Cutoff & maxIPP & 0.5 & Perturb + RF & 0.847 & 0.532 \\ 
  11 & Regression + Cutoff & maxDepth & 0.5 & Perturb + RF & 0.847 & 0.539 \\ 
  11 & Regression + Cutoff & maxIPP & 0.5 & GCFM + XBART & 0.757 & 0.598 \\ 
  11 & Regression + Cutoff & maxDepth & 0.5 & GCFM + XBART & 0.771 & 0.608 \\ 
  11 & Classification & maxDepth & 0.5 & GCFM + Utility & 0.764 & 0.622 \\ 
     \hline
  12 & Regression + Cutoff & maxIPP & 0.5 & Perturb + RF & 0.854 & 0.523 \\ 
  12 & Regression + Cutoff & maxDepth & 0.5 & Perturb + RF & 0.819 & 0.572 \\ 
  12 & Regression + Cutoff & maxIPP & 0.5 & GCFM + XBART & 0.757 & 0.600 \\ 
  12 & Regression + Cutoff & maxDepth & 0.5 & GCFM + XBART & 0.764 & 0.608 \\ 
  12 & Classification & maxDepth & 0.5 & GCFM + Utility & 0.757 & 0.627 \\ 
     \hline
  13 & Regression + Cutoff & maxIPP & 0.5 & Perturb + RF & 0.812 & 0.596 \\ 
  13 & Regression + Cutoff & maxDepth & 0.5 & Perturb + RF & 0.812 & 0.616 \\ 
  13 & Regression + Cutoff & maxIPP & 0.5 & GCFM + XBART & 0.757 & 0.600 \\ 
  13 & Regression + Cutoff & maxDepth & 0.5 & GCFM + XBART & 0.778 & 0.586 \\ 
  13 & Classification & maxDepth & 0.5 & GCFM + Utility & 0.757 & 0.641 \\ 
     \hline
  14 & Regression + Cutoff & maxIPP & 0.5 & Perturb + RF & 0.868 & 0.529 \\ 
  14 & Regression + Cutoff & maxDepth & 0.5 & Perturb + RF & 0.826 & 0.596 \\ 
  14 & Regression + Cutoff & maxIPP & 0.5 & GCFM + XBART & 0.757 & 0.600 \\ 
  14 & Regression + Cutoff & maxDepth & 0.5 & GCFM + XBART & 0.750 & 0.598 \\ 
  14 & Classification & maxDepth & 0.5 & GCFM + Utility & 0.750 & 0.635 \\ 
     \hline
  15 & Regression + Cutoff & maxIPP & 0.5 & Perturb + RF & 0.868 & 0.529 \\ 
  15 & Regression + Cutoff & maxDepth & 0.5 & Perturb + RF & 0.861 & 0.549 \\ 
  15 & Regression + Cutoff & maxIPP & 0.5 & GCFM + XBART & 0.757 & 0.600 \\ 
  15 & Regression + Cutoff & maxDepth & 0.5 & GCFM + XBART & 0.715 & 0.612 \\ 
  15 & Classification & maxDepth & 0.5 & GCFM + Utility & 0.757 & 0.630 \\ 
   \hline
\end{longtable}
  
  \begin{longtable}{ccccccc}
  \hline
Number of Items & Tree Type & Criterion & w & Calibration Data & Sensitivity & Specificity \\ 
  \hline
  2 & Regression + Cutoff & maxIPP & 0.6 & Perturb + RF & 0.924 & 0.364 \\ 
  2 & Regression + Cutoff & maxDepth & 0.6 & Perturb + RF & 1.000 & 0.000 \\ 
  2 & Regression + Cutoff & maxIPP & 0.6 & GCFM + XBART & 0.868 & 0.370 \\ 
  2 & Regression + Cutoff & maxDepth & 0.6 & GCFM + XBART & 0.840 & 0.425 \\ 
  2 & Classification & maxDepth & 0.6 & GCFM + Utility & 0.840 & 0.443 \\ 
     \hline
  3 & Regression + Cutoff & maxIPP & 0.6 & Perturb + RF & 0.938 & 0.326 \\ 
  3 & Regression + Cutoff & maxDepth & 0.6 & Perturb + RF & 1.000 & 0.000 \\ 
  3 & Regression + Cutoff & maxIPP & 0.6 & GCFM + XBART & 0.882 & 0.355 \\ 
  3 & Regression + Cutoff & maxDepth & 0.6 & GCFM + XBART & 0.882 & 0.355 \\ 
  3 & Classification & maxDepth & 0.6 & GCFM + Utility & 0.896 & 0.392 \\ 
     \hline
  4 & Regression + Cutoff & maxIPP & 0.6 & Perturb + RF & 0.979 & 0.250 \\ 
  4 & Regression + Cutoff & maxDepth & 0.6 & Perturb + RF & 0.972 & 0.290 \\ 
  4 & Regression + Cutoff & maxIPP & 0.6 & GCFM + XBART & 0.910 & 0.323 \\ 
  4 & Regression + Cutoff & maxDepth & 0.6 & GCFM + XBART & 0.910 & 0.323 \\ 
  4 & Classification & maxDepth & 0.6 & GCFM + Utility & 0.854 & 0.433 \\ 
     \hline
  5 & Regression + Cutoff & maxIPP & 0.6 & Perturb + RF & 0.986 & 0.212 \\ 
  5 & Regression + Cutoff & maxDepth & 0.6 & Perturb + RF & 0.979 & 0.268 \\ 
  5 & Regression + Cutoff & maxIPP & 0.6 & GCFM + XBART & 0.944 & 0.276 \\ 
  5 & Regression + Cutoff & maxDepth & 0.6 & GCFM + XBART & 0.910 & 0.323 \\ 
  5 & Classification & maxDepth & 0.6 & GCFM + Utility & 0.868 & 0.444 \\ 
     \hline
  6 & Regression + Cutoff & maxIPP & 0.6 & Perturb + RF & 0.986 & 0.206 \\ 
  6 & Regression + Cutoff & maxDepth & 0.6 & Perturb + RF & 0.979 & 0.261 \\ 
  6 & Regression + Cutoff & maxIPP & 0.6 & GCFM + XBART & 0.931 & 0.300 \\ 
  6 & Regression + Cutoff & maxDepth & 0.6 & GCFM + XBART & 0.938 & 0.310 \\ 
  6 & Classification & maxDepth & 0.6 & GCFM + Utility & 0.924 & 0.382 \\ 
     \hline
  7 & Regression + Cutoff & maxIPP & 0.6 & Perturb + RF & 0.972 & 0.266 \\ 
  7 & Regression + Cutoff & maxDepth & 0.6 & Perturb + RF & 0.979 & 0.233 \\ 
  7 & Regression + Cutoff & maxIPP & 0.6 & GCFM + XBART & 0.931 & 0.353 \\ 
  7 & Regression + Cutoff & maxDepth & 0.6 & GCFM + XBART & 0.924 & 0.339 \\ 
  7 & Classification & maxDepth & 0.6 & GCFM + Utility & 0.896 & 0.387 \\ 
     \hline
  8 & Regression + Cutoff & maxIPP & 0.6 & Perturb + RF & 0.972 & 0.285 \\ 
  8 & Regression + Cutoff & maxDepth & 0.6 & Perturb + RF & 0.986 & 0.213 \\ 
  8 & Regression + Cutoff & maxIPP & 0.6 & GCFM + XBART & 0.931 & 0.340 \\ 
  8 & Regression + Cutoff & maxDepth & 0.6 & GCFM + XBART & 0.917 & 0.354 \\ 
  8 & Classification & maxDepth & 0.6 & GCFM + Utility & 0.910 & 0.399 \\ 
     \hline
  9 & Regression + Cutoff & maxIPP & 0.6 & Perturb + RF & 0.972 & 0.312 \\ 
  9 & Regression + Cutoff & maxDepth & 0.6 & Perturb + RF & 0.958 & 0.337 \\ 
  9 & Regression + Cutoff & maxIPP & 0.6 & GCFM + XBART & 0.924 & 0.354 \\ 
  9 & Regression + Cutoff & maxDepth & 0.6 & GCFM + XBART & 0.938 & 0.352 \\ 
  9 & Classification & maxDepth & 0.6 & GCFM + Utility & 0.917 & 0.400 \\ 
     \hline
  10 & Regression + Cutoff & maxIPP & 0.6 & Perturb + RF & 0.979 & 0.278 \\ 
  10 & Regression + Cutoff & maxDepth & 0.6 & Perturb + RF & 0.979 & 0.304 \\ 
  10 & Regression + Cutoff & maxIPP & 0.6 & GCFM + XBART & 0.938 & 0.348 \\ 
  10 & Regression + Cutoff & maxDepth & 0.6 & GCFM + XBART & 0.944 & 0.363 \\ 
  10 & Classification & maxDepth & 0.6 & GCFM + Utility & 0.938 & 0.391 \\ 
     \hline
  11 & Regression + Cutoff & maxIPP & 0.6 & Perturb + RF & 0.979 & 0.266 \\ 
  11 & Regression + Cutoff & maxDepth & 0.6 & Perturb + RF & 0.979 & 0.280 \\ 
  11 & Regression + Cutoff & maxIPP & 0.6 & GCFM + XBART & 0.931 & 0.374 \\ 
  11 & Regression + Cutoff & maxDepth & 0.6 & GCFM + XBART & 0.944 & 0.377 \\ 
  11 & Classification & maxDepth & 0.6 & GCFM + Utility & 0.889 & 0.408 \\ 
     \hline
  12 & Regression + Cutoff & maxIPP & 0.6 & Perturb + RF & 0.979 & 0.271 \\ 
  12 & Regression + Cutoff & maxDepth & 0.6 & Perturb + RF & 0.972 & 0.292 \\ 
  12 & Regression + Cutoff & maxIPP & 0.6 & GCFM + XBART & 0.931 & 0.362 \\ 
  12 & Regression + Cutoff & maxDepth & 0.6 & GCFM + XBART & 0.944 & 0.368 \\ 
  12 & Classification & maxDepth & 0.6 & GCFM + Utility & 0.882 & 0.410 \\ 
     \hline
  13 & Regression + Cutoff & maxIPP & 0.6 & Perturb + RF & 0.986 & 0.260 \\ 
  13 & Regression + Cutoff & maxDepth & 0.6 & Perturb + RF & 0.979 & 0.279 \\ 
  13 & Regression + Cutoff & maxIPP & 0.6 & GCFM + XBART & 0.931 & 0.376 \\ 
  13 & Regression + Cutoff & maxDepth & 0.6 & GCFM + XBART & 0.944 & 0.337 \\ 
  13 & Classification & maxDepth & 0.6 & GCFM + Utility & 0.889 & 0.401 \\ 
     \hline
  14 & Regression + Cutoff & maxIPP & 0.6 & Perturb + RF & 0.986 & 0.262 \\ 
  14 & Regression + Cutoff & maxDepth & 0.6 & Perturb + RF & 0.951 & 0.319 \\ 
  14 & Regression + Cutoff & maxIPP & 0.6 & GCFM + XBART & 0.931 & 0.372 \\ 
  14 & Regression + Cutoff & maxDepth & 0.6 & GCFM + XBART & 0.917 & 0.397 \\ 
  14 & Classification & maxDepth & 0.6 & GCFM + Utility & 0.917 & 0.408 \\ 
     \hline
  15 & Regression + Cutoff & maxIPP & 0.6 & Perturb + RF & 0.986 & 0.256 \\ 
  15 & Regression + Cutoff & maxDepth & 0.6 & Perturb + RF & 0.951 & 0.300 \\ 
  15 & Regression + Cutoff & maxIPP & 0.6 & GCFM + XBART & 0.931 & 0.372 \\ 
  15 & Regression + Cutoff & maxDepth & 0.6 & GCFM + XBART & 0.931 & 0.378 \\ 
  15 & Classification & maxDepth & 0.6 & GCFM + Utility & 0.910 & 0.410 \\ 
   \hline
\end{longtable}

\bibliographystyle{imsart-nameyear} 
\bibliography{bibliography}       